\documentclass[apj,usenatbib,twocolumn]{aastex6}

\usepackage{float}
\usepackage{todonotes}
\usepackage{xspace}
\usepackage{graphicx}
\usepackage{subfigure}
\usepackage{amsmath}

\newcommand{\angstrom}{\textup{\AA}}
\def\kms{{\rm km\,s$^{-1}$}\xspace}
\def\arcsec{$^{\prime\prime}$}

\begin{document} 

   \title{The SED Machine: a robotic spectrograph for fast transient classification}

   \author{Nadejda\ Blagorodnova \altaffilmark{1},
   			James\ D. Neill \altaffilmark{1},
            Richard\ Walters \altaffilmark{1,2}, 
            Shrinivas\ R.\ Kulkarni \altaffilmark{1},
            Christoffer\ Fremling \altaffilmark{1},		
        	Sagi\ Ben-Ami \altaffilmark{3},
            Richard\ G. Dekany \altaffilmark{1,2},
			Jason\ R. Fucik \altaffilmark{2},
			Nick\ Konidaris \altaffilmark{4},
            Reston\ Nash \altaffilmark{2},
            Chow-Choong\  Ngeow \altaffilmark{5},
            Eran\  O.\  Ofek \altaffilmark{6},
            Donal O' Sullivan \altaffilmark{1},
            Robert\  Quimby \altaffilmark{7,8},
			Andreas\  Ritter \altaffilmark{9},
            Karl\  E.\  Vyhmeister \altaffilmark{1,2}}

	\altaffiltext{1}{Cahill Center for Astrophysics, California Institute of Technology, Pasadena, CA 91125, USA }
	\altaffiltext{2}{Caltech Optical Observatories, California Institute of Technology, Pasadena, CA 91125, USA }
	\altaffiltext{3}{Smithsonian Astrophysical Observatory, Harvard-Smithsonian Center for Astrophysics, 60 Garden St., Cambridge, MA 02138, USA}
   \altaffiltext{4}{Observatories of the Carnegie Institution for Science, 813 Santa Barbara St. Pasadena, CA  91101, USA}
    \altaffiltext{5}{ Graduate Institute of Astronomy, National Central University, Jhongli 32001, Taiwan }
	\altaffiltext{6}{ Department of Particle Physics and Astrophysics, Weizmann Institute of Science, Rehovot 7610001, Israel }
	\altaffiltext{7}{Department of Astronomy / Mount Laguna Observatory, San Diego State University, 5500 Campanile Drive, San Diego, CA, 92812-1221}
	\altaffiltext{8}{ Kavli IPMU (WPI), UTIAS, The University of Tokyo, Kashiwa, Chiba 277-8583, Japan}   
\altaffiltext{9}{Department of Astrophysical Sciences, Princeton University, 4 Ivy Lane, New Jersey 08544, USA}


\begin{abstract}
Current time domain facilities are finding several hundreds of transient astronomical events a year. The discovery rate is expected to increase in the future as soon as new surveys such as the Zwicky Transient Facility (ZTF) and the Large Synoptic Sky Survey (LSST) come on line. At the present time, the rate at which transients are classified is approximately one order or magnitude lower than the discovery rate, leading to an increasing ``follow-up drought''. Existing telescopes with moderate aperture can help address this deficit when equipped with spectrographs optimized for spectral classification.  Here, we provide an overview of the design, operations and first results of the Spectral Energy Distribution Machine (SEDM), operating on the Palomar 60-inch telescope (P60). The instrument is optimized for classification and high observing efficiency. It combines a low-resolution (R$\sim$100) integral field unit (IFU) spectrograph with ``Rainbow Camera" (RC), a multi-band field  acquisition camera which also serves as multi-band ($ugri$) photometer. The SEDM was commissioned during the operation of the  intermediate Palomar Transient Factory (iPTF) and has already proved lived up to its promise. The success of the SEDM  demonstrates the value of spectrographs optimized to spectral classification. Introduction of similar spectrographs on existing telescopes will help alleviate the follow-up drought and thereby accelerate the rate of discoveries. 
\end{abstract}


\section{Introduction} \label{sec:intro}

It is now widely agreed that Time Domain Astronomy (TDA) -- the study of transients, variables and moving objects --  is one of the {\it Frontier Fields} of this decade \citep{NAP12951}. The origin of modern time domain astronomy  can be traced to the program begun by  Walter Baade and Fritz Zwicky. The first outcome of this effort was the distinction between novae and \textit{super-novae} \citep{BaadeZwicky1934PNAS}.  In this seminal paper not only the authors remarked on the brilliance of supernovae but also noted the possibility that this phenomena marked the end of a star's life. Two years later Zwicky initiated a dedicated time domain astronomy program with an 18-inch Schmidt telescope survey sited at the Palomar Observatory. By the time of Zwicky's death (1974), the supernovae toll stood at 380 supernovae with 120 found by Zwicky working by himself. The importance of supernovae in building up of the periodic table was a major observational-theoretical achievement of this period \citep{Burbidge1957RvMP}. Coming closer to this era the brilliance of supernovae opened up an entirely new tool for cosmography. Supernovae of type Ia were found to be standardizable candles and led to the first indication of an accelerated expansion of the Universe  \citep{Riess1998AJ,Perlmutter1999ApJ}. So monumental were these two achievements that they were recognized by the Nobel Prize committee. 

The rise of TDA is driven entirely by technological progress in sensors and computing. There are more than a dozen optical TDA surveys, e.g. Catalina Sky Survey \citep{Drake2009ApJ}, PanSTARRS-1 \citep{Kaiser2002SPIE}, the (intermediate) Palomar Transient Factory \cite[iPTF;][]{Rau2009PASP}, ASASSN \citep{Shappee2014ApJ} and ATLAS \citep{Tonry2011PASP}. As a result of these surveys the supernova discovery rate has increased dramatically and the current annual rate of classified supernovae is over a thousand \footnote{Source: http://www.rochesterastronomy.org}.  The large sample size has resulted not only in the discovery of new classes of supernovae (e.g. super-luminous supernovae) but also new sub-classes of extra-galactic explosions (e.g.\ intermediate luminosity infrared transients, luminous red novae,  calcium rich supernovae). Separately, new types of eruptive variables (e.g.\    extreme variability in active galactic nuclei) and  cataclysmic explosions, e.g.\ tidal disruption events (TDEs) are now routinely found. New upcoming surveys such as the Zwicky Transient Facility \citep[ZTF;][]{Bellm2014htu}, BlackGEM \footnote{\url{https://astro.ru.nl/blackgem/}} and Large Synoptic Sky Survey \citep[LSST;][]{ivezic2008lsst} will push these boundaries even further, discovering up to hundreds of new supernovae \textit{every night}.

\begin{figure}[ht]
\centering
\includegraphics[width=0.5\textwidth]{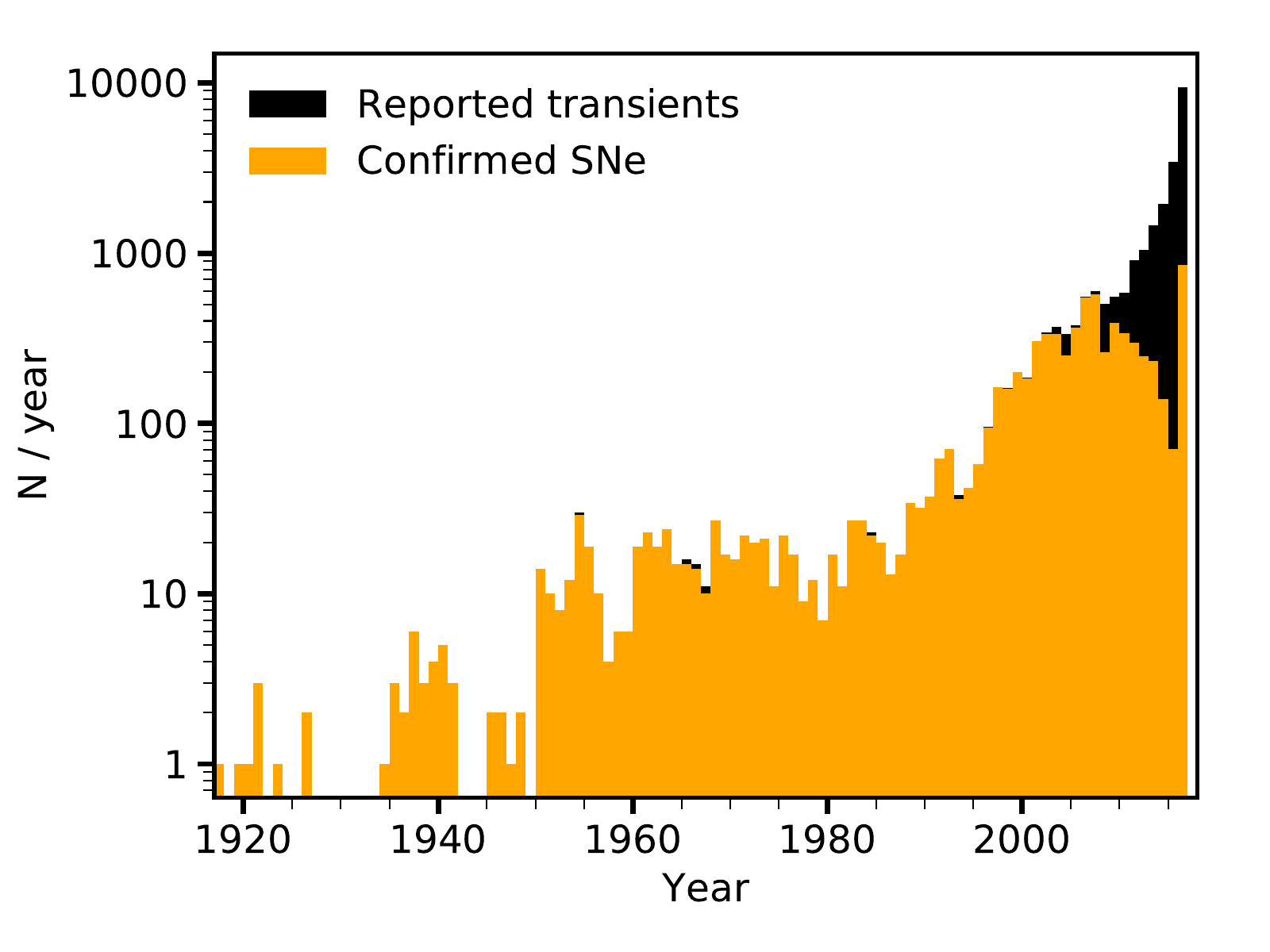}
\caption{ Transient events reported per year (black) and confirmed SNe (orange) per year between 1916 and 2017. Source: \url{http://www.rochesterastronomy.org}. The discoveries may contain a fraction of galactic transients or extreme variability in AGN.  }
\label{fig:sn_discovered}
\end{figure}

In Figure\,\ref{fig:sn_discovered} we display the classified supernova rate. We expect the discovery rate of supernovae to continue a dramatic growth, thanks to the on-going and planned surveys. However, identifying supernovae candidates is merely the first step.  Only with the second step -- spectral classification -- is the discovery process completed. However, spectroscopy requires follow-up observations in most instances on telescopes with apertures larger than that of the discovery telescope. The growing gap between the rate at which transients are identified and classified is truly an ``elephant in the room".

Traditionally low resolution spectroscopy meant a spectral resolution, $R=\lambda/\Delta\lambda\approx 1000$. Fortunately, spectral classification of supernovae and TDEs, with their large expansion velocities, $v/c\approx 1/30$ can be done with lower spectral resolution.\footnote{For a fixed telescope aperture the spectral classification rate increases as $R^{-1}$.}  Next, robotic spectroscopy would be very helpful given the increasing discovery rate of transients. These two considerations have led to ``intermediate resolution'' ($R\approx 400$) spectrographs on robotic telescopes: the Folded Low Order whYte-pupil Double-dispersed Spectrograph (FLOYDS; \citealt{Sand2011AAS}) and SPectrograph for Rapid Acquisition of Transients (SPRAT; \citealt{Piascik2014SPIE}) mounted at the 2-m robotic Liverpool telescope in La Palma (Spain).

The major challenge for robotic spectroscopy is in placing the target precisely on the narrow (1--2\arcsec) slit. This usually results in an iterative target acquisition approach, where the failure rate may increase due to complex host galaxy environment or non-ideal observing conditions. This problem can be solved by an  integral field unit (IFU) spectrograph with a reasonably large field of view (FOV; say $15^{\prime\prime}$ or larger). This approach, apart from the ease of target acquisition, has the additional advantage of the study of the SN environment.

Thus, a major step forward in filling the gap of transient follow-up can be made by deploying a low resolution, robotically controlled IFU spectrograph with $R\approx 100$ and an FOV of, say,  $\approx 30^{\prime\prime}$ square on a dedicated one-meter class telescope with sufficient throughput to achieve a signal-to-noise ratio (SNR) of 5 for a 20.5 $r$-band magnitude transient (the faint end of current TDA surveys) with an exposure time of 3600\,s. The idea of such instrument was conceived in year 2009 as a resource for spectroscopic follow-up of transients discovered by PTF. Here, we present an overview of  ``Spectral Energy Distribution Machine'' (SEDM) optimized for rapid classification spectroscopy. The SEDM is just such an IFU spectrograph and has been in routine operation at the Palomar 60-inch telescope (P60) since 27 April, 2016.

The paper is organized as following. 
Section \ref{sec:overview} provides an overview of the instrument design and main characteristics. Section \ref{sec:operations} describes the daily operations and data flow. The photometric and spectroscopic reduction pipelines are explained in Section \ref{sec:pipeline}. We provide an overview of the performance and a summary of the first scientific results of SEDM in Section \ref{sec:results}. Finally, Section \ref{sec:discussion} offers a discussion on the next steps for the SEDM project and the implications for future transient searches in optical wavelengths.

\section{Instrument Overview} \label{sec:overview}

The SEDM is, by default, the sole instrument for robotic Palomar 60-inch telescope (P60). The telescope optics are Ritchey-Chr\'etien, so that both primary and secondary mirrors have hyperbolic reflecting surfaces. This allows for a wider field of view within a compact design: the tube of the telescope measures 150\,inches (3.8\,m) long. An 18-inch (45.7\,cm) central hole in the primary mirror allows starlight to reach the Cassegrain focus (f/8.75) with a 525-inch (13.3\,m) focal length. The telescope was roboticized in 2004 and single purposed to for imaging photometry, in particular following GRB afterglows \citep{Cenko2006PASP}. In August 2016, the GRB imaging photometer was removed and the SEDM installed at the Cassegrain focus of P60.

\begin{figure}[ht]
\centering
\includegraphics[width=0.5\textwidth]{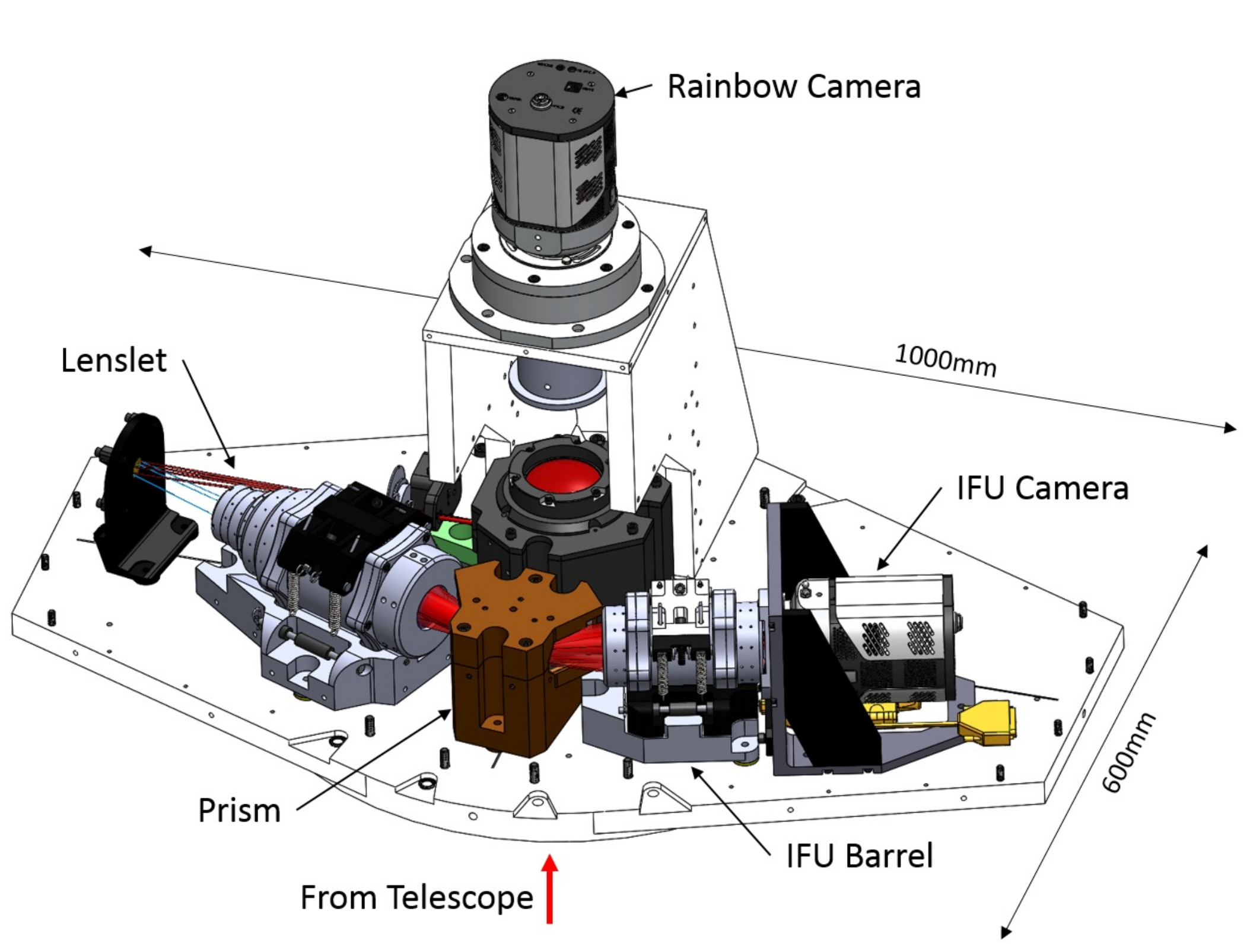}
\caption{Overview of SEDM instrument layout. The light path from the Cassegrain focus is shown with a red arrow. The light for the photometric instrument is collected by the RC, located in the centre. The light for the IFU is initially redirected towards the mirror shown in the left side of the picture and reflected into the lenslet barrel. The light from each lenslet is then dispersed by the triple prism system, placed in the centre. In a final step, the IFU barrel aligns the individual spectra on the detector.  }
\label{fig:layout}
\end{figure}

The SEDM instrument is composed of two main channels: the Rainbow Camera (RC) and the Integral-Field Unit (IFU), mounted in a T-shaped layout (see Figure \ref{fig:layout}). The optics and cameras are protected by an external cover, designed as a sealed container to minimize dust contamination. Overall, the instrument is of a compact size, of 1.0$\times$0.6\,m. The total weight, including electronics, is of $\sim$140\,kg.

The optical path of the instrument is designed so that most of the light from the Cassegrain focus is directed towards the photometric instrument, the RC. The light for the IFU is captured by a pickoff prism that is mounted above the center of the four filter holders for the RC, close to the telescope focal plane. The mirror directs the light through an expander lens to obtain an appropriate scale for the lenslet array (LA). Then, the light path is folded by a flat mirror, and directed through a field-flattening lens onto the lenslet array.  The LA breaks up the field into an array of 2340 hexagonal pupil images that are then sent through a six-element collimator onto the dispersing triple prism.  The triple prism design ensures that the spectroscopic resolution, defined as $R = \Delta\lambda/\lambda$, is at a constant value of $R \approx 100$.  The dispersed pupil images are then focused by a five-element camera through a CCD window that is multi-coated for high transmission. Both the IFU and the RC have relay optics that allow the images to come to focus at their respective CCD detectors, as well as an additional focus mechanism to ensure that the RC and the IFU channels are parfocal. The RC and the IFU are focused by adjusting the position of the movable secondary mirror. The spectrograph is focused using the movable camera optics.

The SEDM  uses two identical Princeton Instruments Pixis  2048B eXelon model 2048$\times$2048 and 13.5$\mu$m pixels CCDs. For each, the CCD window is anti-reflection coated to improve overall detector response.  Thermo-electric coolers keep the detectors at a temperature of $-55^{\circ}$\,C. The excess heat is removed via a glycol cooling system attached to a chiller. A summary of the specifications for the detectors is shown in Table \ref{table:detector}.

\begin{deluxetable}{ll} 
\tablecolumns{2} 
\tablecaption{Specifications for the SEDM detectors.\label{table:detector}} 
\tablehead{ 
\colhead{ Detector feature} & \colhead{Description}  } 
\startdata 
Detector size & 2048$\times$2048 pixels \\
Read noise  (slow) & 5  e$^-$/pix\\
Read noise  (fast) & 22 e$^-$/pix\\
Read time (fast) & 5\,s\\
Read time (slow)  & 40\,s \\
Gain  & 1.8 e$^-$/ADU\\
Pixel size   & 13.5 $\mu$m\\
Saturation  &  55000 ADU \\
Non-linearity &  45000  ADU \\
\enddata 
\end{deluxetable}

\subsection{Rainbow Camera}  \label{sec:rc_instrument}

\begin{figure}[ht]
\hspace{-1.5cm}
\includegraphics[width=1.4\columnwidth]{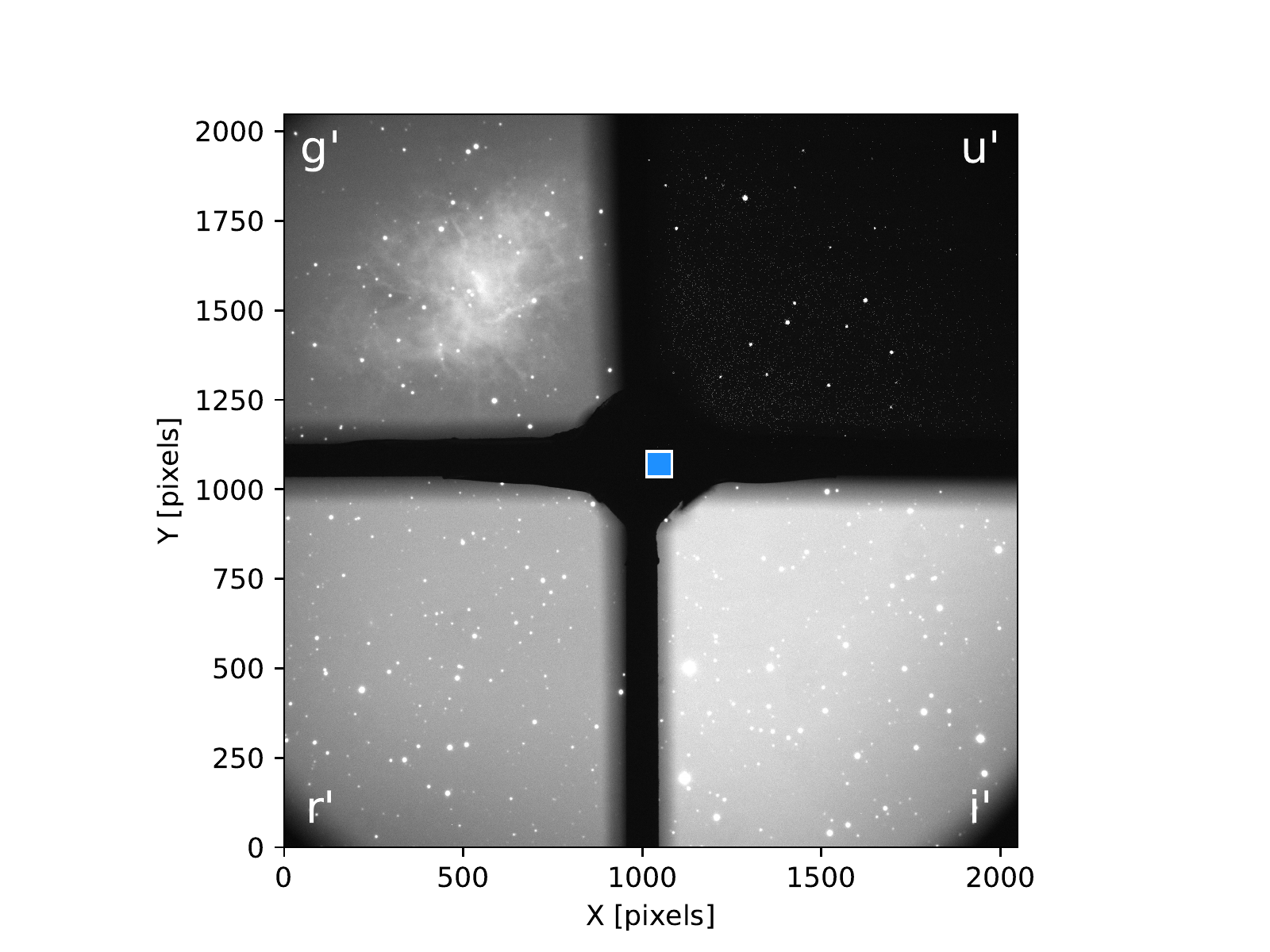}
\caption{Overview of an RC raw image, obtained at the Palomar 60-inch telescope. The Crab Nebula can be seen in the upper left corner, corresponding to the $g$-band. As it can be appreciated in the image, the background level for each filter is different, with the u-band being the least sensitive. Light reaching at the center of the cross is directed towards the IFU, the FOV of which FOV is shown as a blue rectangle.}
\label{fig:rcimage}
\end{figure}

The RC is used for guiding, calibration, target acquisition and science imaging. The 13$\times$13\,arcmin FOV is split into four quadrants, one for each filter: $u'$, $g'$, $r'$ and $i'$. Because part of the field is obscured by the filter holder, only $\sim 6\times$6\,arcmin field is available for imaging the target with one filter. In order to obtain an exposure with a different filter, the telescope is offset to center the target in a new quadrant. A raw SEDM image is shown in Figure \ref{fig:rcimage}, where the Crab Nebula can be seen in the $g'$ panel. The blue rectangle shows the position for the IFU.  

\vspace{-2cm}

\begin{figure}[ht]
\begin{center}
\centering
\includegraphics[width=0.5\textwidth]{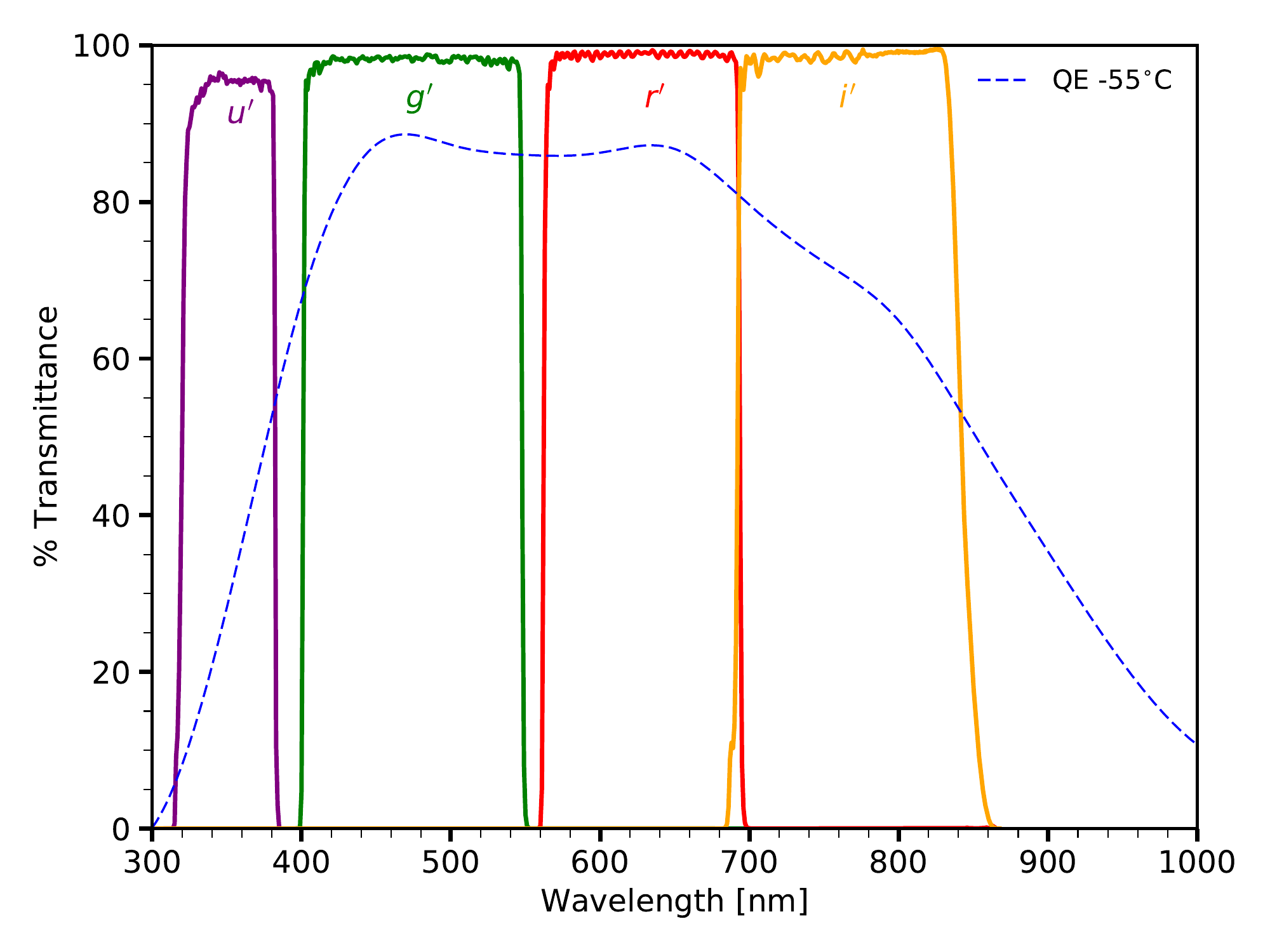}
\end{center}
\caption{ \textit{Astrodon} generation 2 filters transmittance used for the RC. The CCD quantum efficiency at current operating temperature is shown with a dashed line. }
\label{fig:filters}
\end{figure}

The RC uses \textit{Astrodon}\footnote{\url{http://www.astrodon.com}} generation 2 filters. Their transmittance are shown in Figure \ref{fig:filters}, along with the predicted quantum efficiency of the CCDs at the current operating temperature. The specifications for each filter, pixel size and single filter FOV are shown in the upper part of Table \ref{table:instrument}.

\subsection{Integral Field Unit} \label{sec:ifu_instrument}

\begin{figure}[h]
\begin{center}
\centering
\includegraphics[width=1\columnwidth]{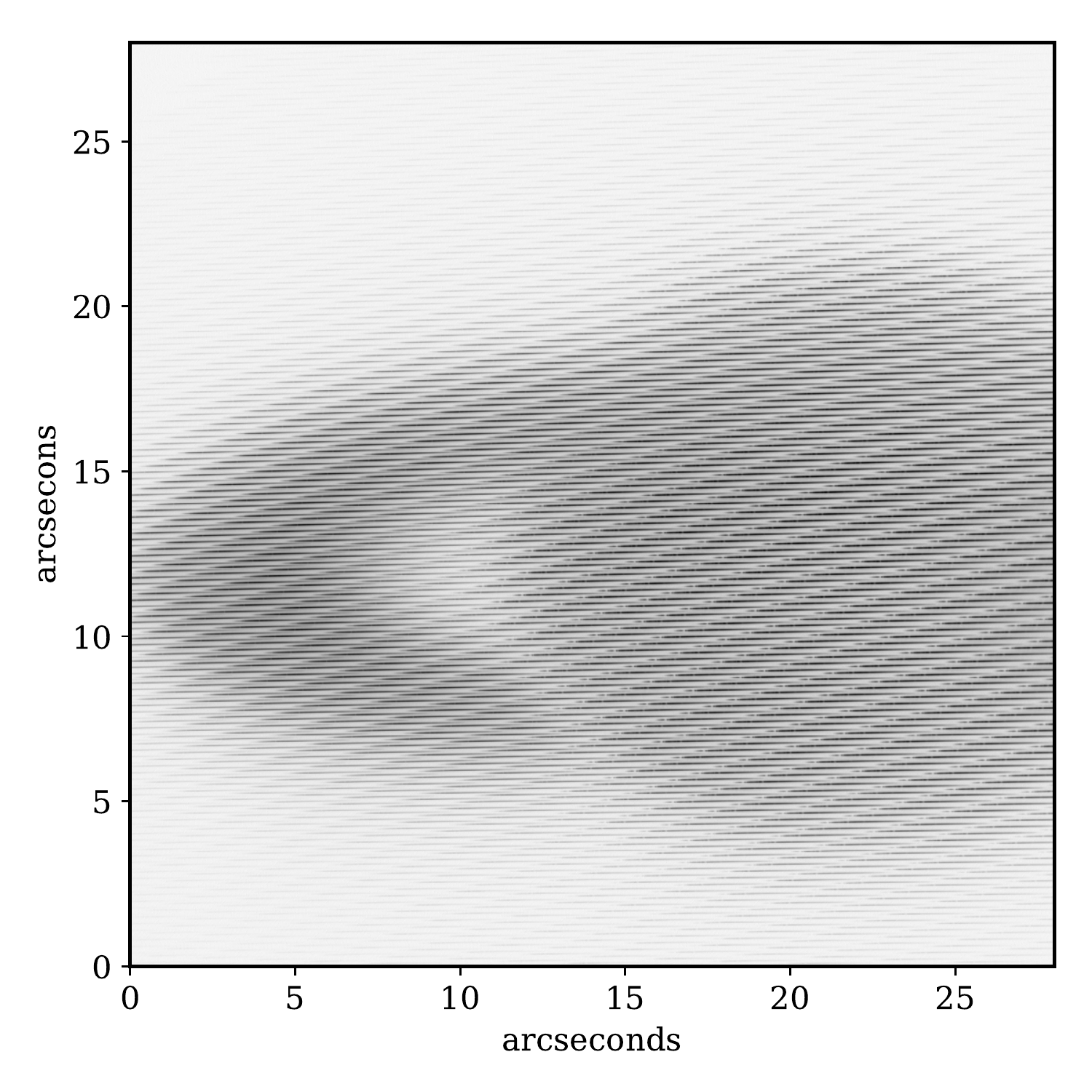}
\end{center}
\caption{Raw negative IFU image of Saturn. The exposure time is 1\,s. The size of the IFU covers 28$\times$28 \arcsec.}
\label{fig:saturn}
\end{figure}

The SEDM IFU is a lenslet array covering $28\times28$\,\arcsec with 52$\times$45 hexagonal lenslets that project their beams through a triple prism that produces a constant spectral resolution of $R
= \Delta\lambda/\lambda \sim 100$.  
A pickoff prism intercepts a field of size $28\times28$\,\arcsec  from the center of the {\it f}/8.75 P60 beam. The pickoff  location minimizes the field lost by the RC.  The IFU beam relayed by the pickoff prism is magnified by a doublet expander lens and field-flattened by a singlet lens onto the lenslet array via a flat mirror that folds the light path. The lenslet array divides the relayed beam into 2340 pupils of 0.75\,\arcsec in diameter, which are then collimated onto the triple prism.  The dispersed beams that emerge from the triple prism are focused by the camera optics onto the CCD. As an example, Figure \ref{fig:saturn} shows the IFU image of Saturn as directly projected on the detector. A summary of the specifications of the instrument are shown in lower panel of Table \ref{table:instrument}.

\begin{table}
\begin{tabular}{ll} \hline
\tabletypesize{\footnotesize} 
RC specification & Descriptions\\ \hline
FOV & 6.0$\times$6.0 arcmin  \\
Pixel scale & 0.394\arcsec \\
$\lambda_{\rm{eff}} (u')$ & 355\,nm \\ 
$\lambda_{\rm{eff}} (g')$ & 477\,nm \\
$\lambda_{\rm{eff}} (r')$ & 623\,nm \\
$\lambda_{\rm{eff}} (i')$ & 762\,nm  \\
$\Delta \lambda (u')$ & 67 \,nm \\ 
$\Delta \lambda (g')$ & 148 \,nm \\
$\Delta \lambda (r')$ & 134 \,nm \\
$\Delta \lambda (i')$ & 170 \,nm  \\ \hline
IFU specification & Description \\ \hline
FOV & 28$\times$28 \arcsec  \\
Pixel size & 0.125 \arcsec \\
Lenslet array size (spaxels) & 45 $\times$ 52\\ 
Single lenslet diameter &  0.75\arcsec \\
Lenslet size & 52$\times$45 \\
R & $\sim$100 \\
Dispersion &  17.4$-$35 \angstrom /pix\\ \hline
\end{tabular}
\caption{Instrument specifications for the RC and IFU.\label{table:instrument}}
\end{table}


\section{Operations and data flow} \label{sec:operations}

\begin{figure}[h]
\begin{center}
\centering
\includegraphics[width=0.5\textwidth]{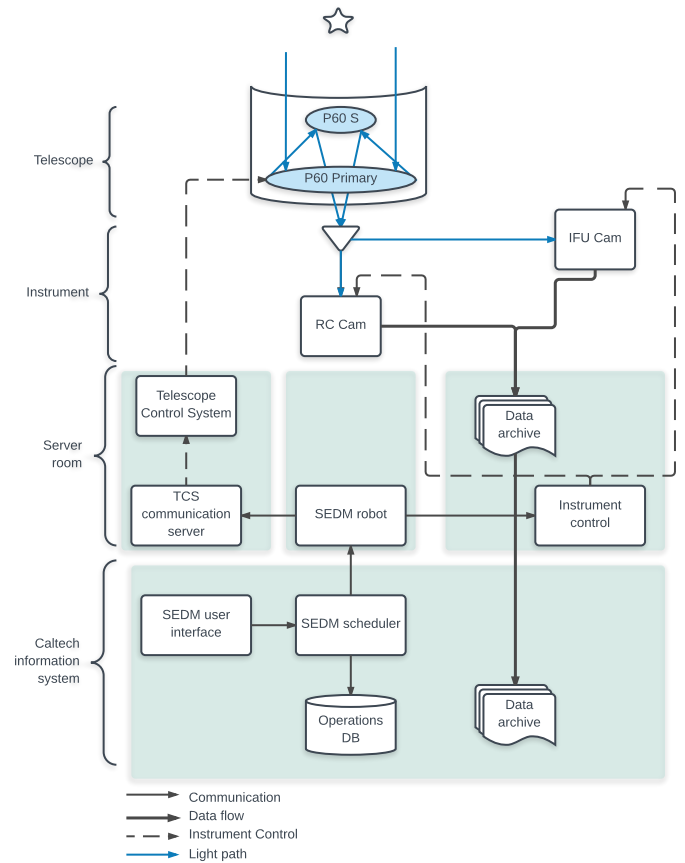}
\end{center}
\caption{ Schematic representation of the SEDM hardware and software modules. A web user interface allows to create observation requests, which are scheduled for the night. The robot uses the schedule to communicate with the telescope and the instrument to carry out the exposures and send the data from Palomar to the Caltech data archive. Each machine is shown as a green box. }
\label{fig:dataflow}
\end{figure}

Figure \ref{fig:dataflow} provides a high level graphical summary of the SEDM observing system. 

\subsection{SEDM Startup}
What follows below is a description of the flow of activities for a typical day. At 1600 Pacific time the Palomar crew release P60 to the astronomical operations group. The master program then initializes the main software module controlling the instrument (the ``SEDM robot") and also creates a simulated schedule for the night. Diagnostic checks are undertaken to ensure that the various sub-systems (telescope, dome, IFU \& RC, data processing computer and archival computer).  Should a failure be detected then an automated warning is issued. On-site personnel then (typically) have about two hours to diagnose and fix the problem. 

Once all diagnostic checks are successful the automation script begins a series of bias, arc-lamps, and dome flats using the IFU camera. For the RC camera, in addition to the above, a set of twilight flats are undertaken (provided the weather conditions permit opening of the dome).

Just before nautical twilight, the system runs a focus loop using the secondary stage on the Palomar 60-inch.  A bright star is selected from the Smithsonian Astrophysical Observatory (SAO) star catalog according to visibility and airmass constraints. The focus for the RC is obtained from a set of short exposures on the target. The last best focus position is used as an initial position, and the stage of the secondary mirror moves 5 steps on each side, with an increment of 0.25\,mm. A two degree polynomial is fit to the FWHM of the images to interpolate the new best focus position. An similar process is followed for finding the focus of the IFU, but in this case we fit the PSF of the image obtained by collapsing 3-D IFU cube along the wavelength axis.

At the beginning of nautical twilight, the robot takes a standard star observation before acquiring the first science target.

\subsection{Acquisition process}

\begin{figure}[h]
\begin{center}
\centering
\includegraphics[width=0.5\textwidth]{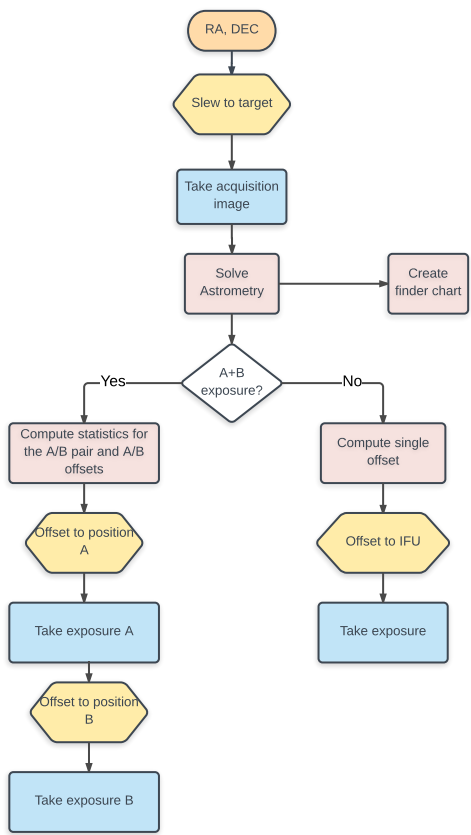}
\end{center}
\caption{ Acquisition process followed by the robot. The steps performed by the telescope are encoded in yellow whereas the instrument actions are shown in blue. In red, we show the steps executed by the software acquisition modules.}
\label{fig:acquisition}
\end{figure}

The acquisition process for IFU science targets is shown in Figure \ref{fig:acquisition}. For bright targets (\,$<$16.5\,mag) only one exposure is undertaken (hereafter ``A" exposure). For stars fainter than this magnitude two exposures with identical exposure time but with an angular offset are undertaken (``A" and ``B"). The paired images allow for a simple and robust  subtraction of the background. In order to accurately compute the offset for the target, a 10--30\,s acquisition exposure is undertaken with the RC.  The resultant image is then analyzed using a custom astrometry fitting program to obtain the true pointing of the telescope.  Typically, this step takes 20--30\,s depending on the field. The robot uses this information to accurately offset the telescope, so that the target falls close to  center of the field of view of the IFU. In the case of a combined A and B exposures, the system also analyzes the statistical properties of the background where the B pair, relative to A, is going to be placed. The goal is to choose the emptiest location for the pair, so that with a simple subtraction operation will result in efficient removal of the sky. Following the last exposure is complete, the robot requests the slew of the telescope to the next field. During this period the CCD is read. This way the on-sky efficiency is maximized. All acquisition images are used to create finder charts for the objects. These charts are helpful during the IFU data reduction.

Under normal circumstances, the system operates autonomously through the night. The astronomical observations end with flat-field exposures undertaken at morning 
nautical twilight. The telescope is then stowed and a summary email of the night's observations is sent out. 

Errors that occur during the night are handled and logged by the automation script.  In the event of an unrecoverable error, the system is programmed to stow the telescope, close the dome, and send out an email and text alert indicating the problem.

\subsection{SEDM Observatory Control System}

The SEDM Observatory Control System (SOCS) is a custom built \texttt{python} code designed to integrate hardware components, scheduling and processing software into one program.  The \texttt{python} module incorporates six sub packages that provide access to various components. By compartmentalizing these functions, the system can be easily modified and expanded in the future. The  functions carried out by the packages are described below.

\begin{itemize}
\item \textbf{Camera:} This package handle communications with the cameras. The RC and IFU cameras are off-the-shelf Princeton Pixis cameras (E2V 2048$\times$2048: CCD42-40).  Communication with these cameras are established via fiber to USB connection and a custom  \texttt{python} package that allows the setting of various camera parameters (exposure time, gain, readout speed
amplifier, and shutter).  The drivers for these cameras are currently configured for \texttt{Windows}, so that they run in a separate machine.  The SOCS script communicates with this \texttt{Windows} machine via a TCP/IP socket connection.  

\item \textbf{Observatory:}
This package contains scripts to handle all the physical components -- telescope, dome and arc lamps -- of the observatory.    Communications with the telescope and dome systems occur via a TCP/IP socket connection to the Telescope Control System (TCS).  The arc-lamps are controlled via a network power switch.

\item \textbf{Sky:}  
The sky package contains scripts for calculating the observing times for a night and a special targets class.  Observing times are calculated using the \texttt{pyephem} package.  A custom \texttt{python} class was written to create target objects.  These objects contain all the meta data for each requested science target observation.  This information includes the objects RA, DEC, observing constraints, and history of the observations already taken.  Targets are complied from a \texttt{SQL} database which gets updated in real time as the observations take place.  

\item \textbf{Scheduler:}
All targets are assigned a numerical priority number typically ranging from 1-5.  The scheduling module then rearranges targets based on input constraints for observing cadence, Moon separation, Moon illumination, and airmass constraint.  A printout of the observing plan for the night is then made.  For the rest of the night, any new targets assigned in real time into the queue will trigger the rescheduling of the night's plan.

\item \textbf{Sanity:}
The \texttt{sanity} package serves to make sure the telescope and cameras are all functioning properly and to prevent any actions that may cause damage to the equipment.  Examples include ensuring a science target is not below the horizon limit, that the telescope will not track into a limit during a long exposure, and checking that the cameras are cooled.  In addition, the \texttt{sanity} package also contains the error handling routines.

\item \textbf{Utilities:}
The \texttt{utilities} package contains tools that help with the normal operations of the robot.  These tools include \texttt{FITS} header management, World Coordinate System (WCS) solving, extracting best focus, and picking the best orientation for A/B pair observations using the IFU.  

\end{itemize}

\subsection{Guider}

The tracking/guider code for the SEDM is written in \texttt{python} and uses Telnet to communicate with the 60-inch telescope guider, automatically issuing adjustments when the telescope drifts from its target. The code is stand-alone and uses only standard \texttt{Python} packages with the exception of \texttt{telnetlib} for communication. As such it can be readily adapted for other uses. First, a telnet connection is established with the telescope guider via hard-coded network parameters in the main configuration file. The program then continually searches for new images in the instrument's data directory, downloading them when available. Once an image is downloaded, the header is checked to see if any intentional moves or target changes have occurred. If a target change or intentional move has occurred, the tracking parameters are reset and a new guide-star is automatically located. The latter is done through a custom star-finding algorithm written into the accompanying package, \texttt{sedmtools} This algorithm searches for the brightest pixels in the image below a certain cap value which it considers to be saturated. If a saturated pixel is found, an iterative method is initiated which incrementally grows a rectangular mask to cover the saturated region and thus prevent it from leading to further false detections. Once a usable star is found, the coordinates of its centroid are returned to the main method. A small box is then defined in image coordinates and its position/size remain fixed until another intentional move is made. For each new image until then, the same box is extracted around the guide star and compared to the previous image via a 2D correlation matrix, the location of the peak of which indicates any offset or shift between the two images. For sub-pixel accuracy, a 2-dimensional Gaussian is fit to the correlation matrix and its peak used as the overlap between the two images. The X and Y offsets in image coordinates are converted into a change in RA and Dec, and a command is sent to the guider to nudge the telescope pointing in the opposite direction. Finally, the X and Y FWHM of the guide star is measured and logged for every image to keep track of seeing and focus conditions.

\section{Pipelines} \label{sec:pipeline}

\subsection{Photometric data reduction pipeline}
 \label{sec:rc_pipeline}

The photometric pipeline reduces the imaging data taken with the RC. In this section, we explain the \texttt{python} based photometric reduction pipeline and calibration of the zeropoint. We describe the SEDM aperture photometry and difference imaging photometric pipelines, used to build the lightcurves for the observed objects.

\begin{figure}[h]
\begin{center}
\centering
\includegraphics[width=0.5\textwidth]{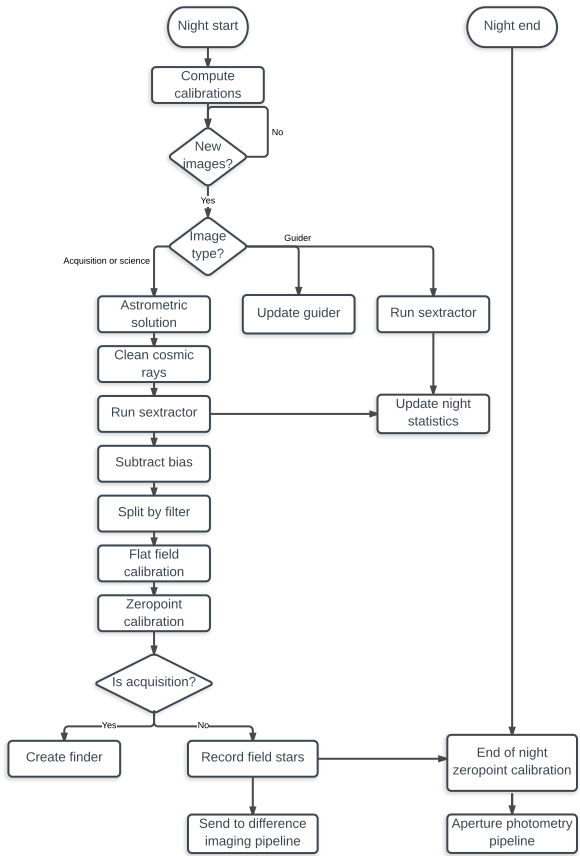}
\end{center}
\caption{Data flow for the nightly data reduction pipeline. The beginning-of-night trigger activates the on-the-fly data reduction and sends the images to the difference imaging pipeline. The end-of-night trigger starts the final zeropoint calibration using all the data taken during the night, and concludes with the photometric data reduction process using the aperture photometry pipeline.}
\label{fig:photpipe}
\end{figure}

\subsubsection{Calibrations}
The photometric pipeline uses standard routines for calibrations. The bias frames, taken as part of the afternoon calibrations, are median-combined into a master bias, which is subtracted from each science frame. We use the \textsc{IRAF} routine \texttt{imcombine} selecting the median value for the stacked image. Two different master bias are created, one for each read-out mode.

The flat-field calibration uses sky flats only, as the dome flats are not representative enough of the light projection on the focal plane: when dome flats are used, the structure that holds the filters creates a shadow effect on the focal plane. This effect, which is amplified at the edges of the FOV, is not representative of the light distribution for on-sky exposures. The problem is naturally solved by using twilight flats. Given the quantum efficiency of the CCD and the filter transmittance, each filter requires different exposure time to reach the optimum number of counts ($\sim$30,000). Given the limited twilight time, the flat fields are obtained with a range of exposure times, favoring one filter at a time. After slicing the RC exposure into four different files, one for each filter, the master flat for each filter is created by subtracting the master bias and computing the median for all exposures with counts in the linear regime (between 15,000 and 45,000). In case the calibrations were not taken because of atmospheric conditions, the pipeline uses the flat fields created during the previous night of operations.

\subsubsection{Reduction process}
Images produced by the RC contain all four filters, as shown in Figure \ref{fig:rcimage}. At a time, the object of interest only falls into one of the quadrants, as shown for the Crab Nebula in this Figure. The same exposure time produces different levels of background, and therefore each filter needs to be processed separately. However, as a first step, the pipeline runs \textsc{Astrometry.net} \citep{Lang2010} on the complete raw image in order to obtain a WCS solution using all filters at the same time. Because of the low sensitivity of the $u$-band, short exposure times would not allow to generate a reliable astrometric calibration for this filter only.  Once the WCS is registered, the raw image with the astrometric solution is divided into four sub-images or slices, containing the data for each one of the filters. Although the target falls into only one quadrant at a time, all four slices are reduced in order to improve the zero-point calculation. 

In the next step, we run the cosmic ray rejection code, which detects and interpolates the pixels affected by cosmic rays. We use the \texttt{python} module \texttt{cosmics} \citep{vanDokkum2001PASP}. The gain and read-out-noise for the read (fast/slow) are used. As a threshold, an 8$\sigma$ level is used.

As part of the reduction process, also \texttt{SExtractor} \citep{BertinArnouts1996AA} is run on each image in order to asses the total number of sources in the image, the FWHM of the exposure, the average ellipticity of the sources and the level of background. This information is used to monitor the focus of the instrument and the overall quality of the image. A clouded exposure will produce a low number of extracted sources.

\subsubsection{Zeropoint calibration}

The SEDM photometry is based on the AB system \citep{OkeGunn1983}, used by the Sloan Digital Sky Survey (SDSS) \citep{Fukugita1996} and the PanSTARRS \citep{Magnier2016arXiv} photometric systems. 

The zeropoints for SEDM are computed in two different ways: per image basis and per night basis. When a single image is used, the central field coordinates are used to query the SDSS/PS1 catalogues, in order to extract all stars with magnitudes 13.5$-$21.0\,mag for $g,r,i$ calibration, and  13.5$-$19.0\,mag for $u$-band calibration. Two different filters are applied to these set of stars to reduce possible contamination. We discard stars in the pixels close to the edge of the frame. Each candidate field star is checked for having bright neighbors (magnitudes $<$\,20.0 or $<$\,19 for $u$-band) within 15\arcsec. Only isolated stars are selected. Their positions are used to test their detection in the science frame. For each star, if there is a detection with a FWHM comparable to the value derived from \texttt{SExtractor}, then the star is added to the zeropoint calibration set. The instrumental magnitudes for the set is obtained running aperture photometry (see Section \ref{sec:app_phot}). Finally, the zeropoint is computed by fitting a one degree polynomial to the instrumental and catalog magnitudes. In the X-axis, we consider the color of the source: $u-g$ for the $u$-band, $g-r$ for $g$-band, $r-i$ for $r$-band and $i-r$ for $i$-band. In the Y-axis, we measure the difference between the catalog and the instrumental magnitude. The catalog and measurement errors are added in quadrature and used as weights in the fitting procedure. The fitting is done in two consecutive steps. In the first one, the data are fit to determine the coefficients of the polynomial. After that, we use these results to predict the zeropoint for all the field stars, and compute the median absolute deviation (MAD) as a robust statistic to quantify the scatter in the fit. In the second step, we discard all measurements with errors in predicted magnitudes larger than 3$\times$MAD, so that we minimize the impact of outliers. The polynomial coefficients from the second fit are saved as the image zeropoint and color term for the image.

At the end of the night, a global fit is run to obtain a nightly solution. Our least square minimization considers the following variables: 1) the colour of the source, 2) the average airmass of the exposure minus $1.3$ (default in SDSS calibrations), 3) the time of the exposure as quadratic dependence, 4) the X,Y position on the CCD as a quadratic dependence and 5) the FWHM for the exposure.

The global least square equation for the zeropoint $zp$, can be expressed as:

\begin{equation}
zp = zp_0 + a C + b (\zeta-1.3)+ M_{cal}\\
\end{equation}

where $zp_0$ is the nominal, colour independent zeropoint for a given filter, $C$ is the colour index term ($u-g$, $g-r$, $r-i$ or $i-r$), $\zeta$ represents the Hardie airmass and $M_{cal}$ is the additional calibration term accounting for time of the exposure ($t$), location on the CCD ($X$ and $Y$) and the FWHM of the exposure.

\begin{equation}
M_{cal}=c\,t + d t^2 + e X + f Y + g X^2 + h Y^2 + j FWHM
\end{equation}

The coefficients $a,b,c,d,e,f,g,h,j$ are computed using the least square method. Similar to our single image zeropoint, we use two iterations in order to reduce the number of outliers and obtain a more reliable fit.

The coefficients from this global solution are recorded and used as a higher quality zeropoint for photometric nights. The RMS between the predicted and the measured values is included as the error in the zeropoint estimation.

\subsubsection{Aperture photometry pipeline}\label{sec:app_phot}

The aperture photometry pipeline is based on the \texttt{IRAF} routine \texttt{apphot}. In order to adapt the aperture to the seeing conditions of the exposure, we use the FWHM derived from \texttt{SExtractor} measurements. The aperture radius is selected to be 2$\times$FWHM, so that most part of the flux is collected. The inner sky radius is set to four times the aperture radius with a width of 15 pixels. The instrumental magnitude of the source, the uncertainty, and the sky background are recorded in the image header and stored in the database for that particular observation.

The pipeline also generates a summary file with all the science observations, along with their instrumental and calibrated magnitudes.

\subsubsection{Difference imaging pipeline}

To produce host-galaxy subtracted photometry the bias-subtracted and flat-fielded science images are processed with an adaptation of the Fremling Automated pipeline (\texttt{FPipe}; see \citealp{2016A&A...593A..68F} for a detailed description of the processing steps performed by the pipeline). Our SEDM RC adaptation of \texttt{FPipe} contains only minor modifications, with the largest being in the image registration step. The pipeline starts by constructing reference images from mosaiced SDSS frames\footnote{The pipeline can also utilize PS1 image stacks as references. However, the PS1 filters are not as well matched to the filters mounted on the SEDM RC, resulting in considerably stronger residuals for strong point-sources compared to when SDSS references are used.}, by stacking them using \texttt{SWarp}, to make sure that the entire field that was observed in each science frame by the RC is covered. The science frames are then registered to the reference mosaics using transformations of increasing complexity depending on the number of common point sources found in the frames. This starts from just a rotation, scaling and shifting for less than 7 common sources, and ends with a third order polynomial transformation for above 25 common sources, which we find necessary in some cases, to achieve an RMS of $\sim0.1$ pixels in the registrations. The detection threshold used in \texttt{SExtractor} is also progressively increased, depending on the number of common sources identified, to maximize the amount of common sources detected while also minimizing the amount of false-detections by \texttt{SExtractor}. The registration procedure described above is done in an iterative process. We go from a low to high detection threshold and allow up to 6 variations of our registrations to be performed. The registration with the best combination of RMS and number of common point-sources is then identified and used.

After image registration, the science frames are PSF matched to the references using the Common PSF Method (CPM) \citep{Gal-Yam2008ApJ}. We model the PSF in the reference frame and in each science frame using  \texttt{SExtractor} in combination with  \texttt{PSFex} \citep{Bertin2011ASPC}. Assuming that the PSF is constant across the CCD, we clean and stack isolated sources, measuring the PSF from the raw data. We use this model to perform the photometry on common SDSS or PS1 stars, to find the zeropoints of both frames. These measurements are then used to put the science and reference frames on an equal intensity scale, after which the reference image is subtracted from the science frame, and the final photometry is measured via a PSF fit on the transient. Limits and uncertainties are determined via insertion of artificial sources. Detections versus non-detections are determined from the quality of the fit of the PSF model to the transient (see again \citealp{2016A&A...593A..68F}, for details). For our SEDM adaptation of \texttt{FPipe} we have also added a simulation of the potential error due to the image registrations. This is done by re-doing the subtraction and photometry process after the registration step for small shifts of the reference image with respect to the science frame, with the size of the shifts corresponding to the RMS of the registrations. The contribution from the registration uncertainty to the transient flux is then taken as the standard deviation of these results.

The turnaround time for one photometric measurement for the SEDM RC using \texttt{FPipe} is on the order of 90\,s. Most of the processing time is consumed by the iterative registration process.

\begin{figure}[h]
\begin{center}
\centering
\includegraphics[width=0.48\textwidth]{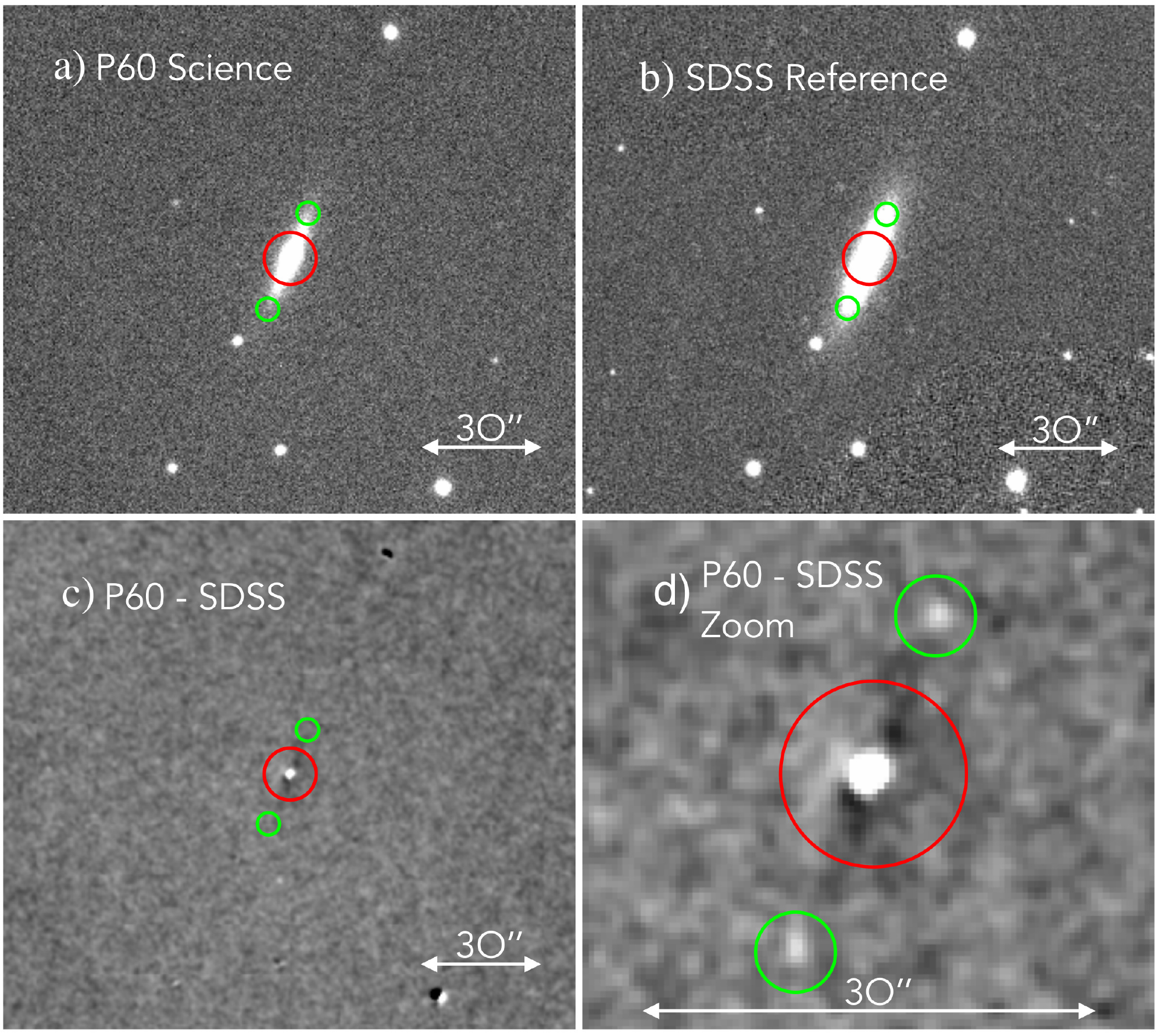}
\end{center}
\caption{Example on a difference imaging follow-up photometry for a nuclear transient using \texttt{FPipe} for SEDM on the P60 telescope. We use the TDE iPTF16fnl \citep{Blagorodnova2017ApJ} as an example.  
Panel \textit{a)} shows the first P60 $g$-band follow-up image from the RC. The position of the transient is indicated by a red circle.
Panel \textit{b)} shows the corresponding SDSS reference image. Panel \textit{c)} shows the resulting difference image between the RC exposure and the SDSS image after registration, PSF and ZP matching. In Panel \textit{d)} we show the result of inserting two simulated sources with $g$-band magnitude of 22\,mag in the image shown in Panel \textit{a)} at the position of the green circles and redoing the subtraction procedure.
}
\label{fig:diffpipeline}
\end{figure}

Using SDSS references and typical exposure times, a transient source can be followed until it fades below $\sim$22\,mag, if in the outskirts of an extended galaxy, and to 21\,mag in the center of a galaxy core (although this also depends on the brightness of the galaxy core). As an example, in Figure~\ref{fig:diffpipeline} we show a reduction in the $g$-band of the nuclear transient iPTF16fnl \citep{Blagorodnova2017ApJ}. This event was located at the center of a bright host (15.49\,mag in the $g$-band in SDSS) and it was successfully followed to 20.3\,mag using the SEDM with exposures $<300$\,s. In the science frame shown in Figure~\ref{fig:diffpipeline} the transient had a $g$-band magnitude of 19.6, and a SNR of above 50 at the signal peak in the difference image. We also show a simulation of the depth that can be reached in the outskirts of this galaxy by inserting artificial sources in the science frame and re-running \texttt{FPipe} (see Panel \textit{d} of Figure~\ref{fig:diffpipeline}). The SNR at peak of these artificial sources is around 5.

\subsection{Integral Field Unit} \label{sec:ifu_pipeline}

\begin{figure}[h]
\begin{center}
\centering
\includegraphics[width=0.5\textwidth]{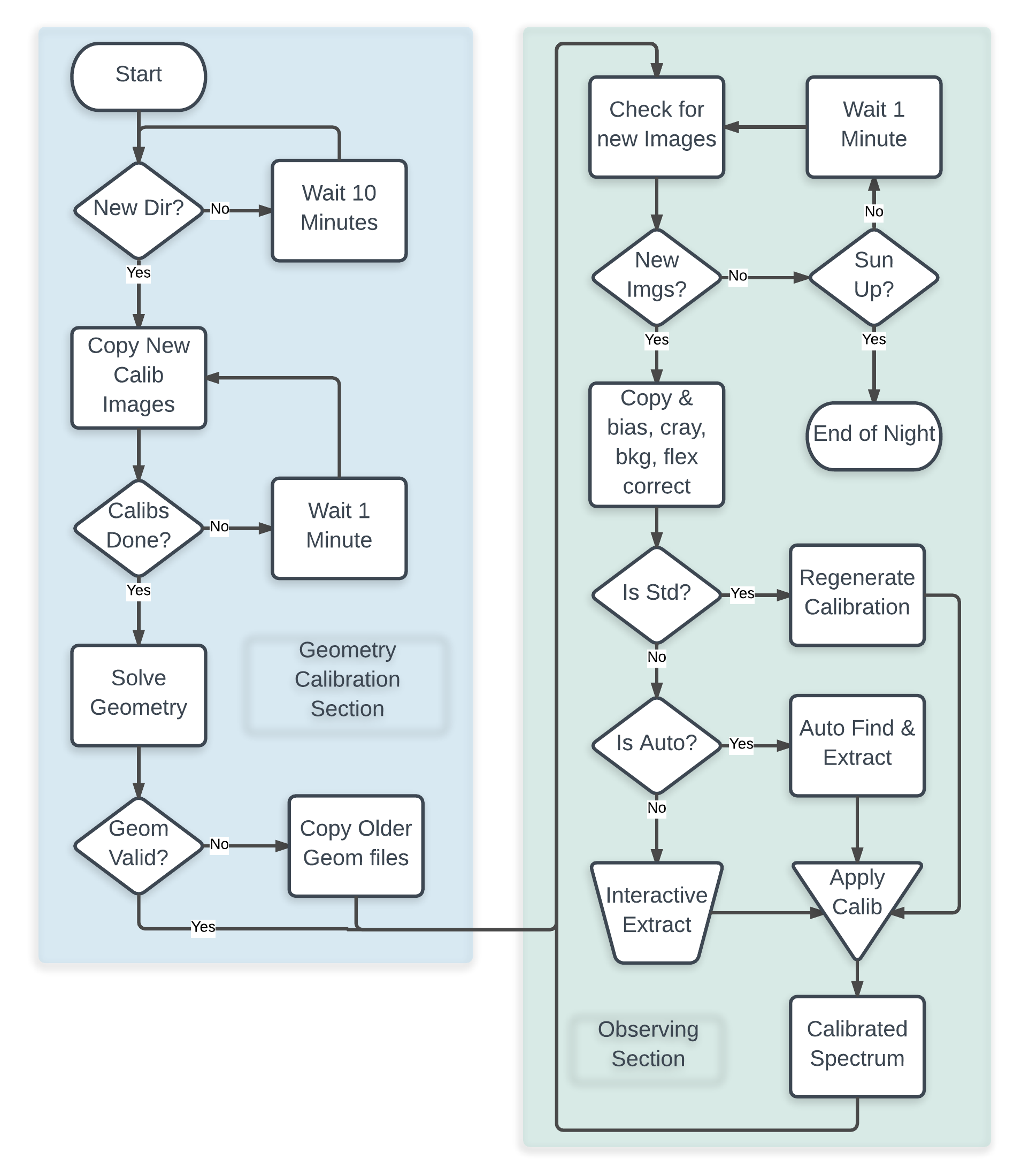}
\end{center}
\caption{ Data flow for the nightly IFU data reduction pipeline.  The geometry calibration section is done late in the afternoon, while the observing section is done after sunset.}
\label{fig:ifudrp}
\end{figure}

The SEDM IFU data reduction pipeline (DRP) operates automatically to reduce the images taken by the IFU camera.  The processing flow is shown in Figure \ref{fig:ifudrp}. The processing consists of basic image reduction, then defining the IFU spatial and wavelength geometry followed by the spectral reduction and extraction stage.  Extracted spectra are then calibrated using an inverse sensitivity curve that is generated from spectrophotometric standard star observations.

An alternative pipeline for SEDM was developed in the \texttt{MATLAB} environment and is available as part of the \texttt{MATLAB} Astronomy \& Astrophysics Toolbox\footnote{https://webhome.weizmann.ac.il/home/eofek/matlab/} \citep{Ofek2014ascl}. This pipeline was used mainly for studying how to reduce the SEDM data given the challenges in wavelength calibration and background
subtraction. The following describes the \texttt{python} pipeline.

\subsubsection{Basic Reduction} 

The basic calibration set is taken during the late afternoon and consists of a set of bias frames, two sets of dome flats illuminated by a halogen lamp and by a Xe lamp, and a set of internal arcs illuminated by Hg and Cd arc lamps.  All these calibration images are taken with the telescope at the same position to minimize flexure offsets.  A master bias image is generated from the ten bias frames of zero exposure time.  This master bias frame is then subtracted from all images. This is followed by a cosmic ray rejection stage that uses the \texttt{lacosmicx}\footnote{github.com/cmccully/lacosmicx} \texttt{python} package.  This package runs using \texttt{cython}, which makes this step faster than other cosmic ray rejection routines that run in native \texttt{python}.  We also tune the parameters such that the peaks of the spaxels are not rejected by using the \texttt{gaussy} PSF model that assumes that real objects have a Gaussian shape in the y-direction, but are extended in the x-direction.  Once all the required calibration frames are reduced, the spatial and wavelength geometry can be defined.

\subsubsection{Geometry Definition} The geometry definition stage uses a set of seven dome-flat images to define the spatial location of the spectral traces in the CCD pixel coordinate system.  Arc-lamp image sets of Hg, Cd, and Xe arcs (seven of each) are used to define the wavelength solution for each individual trace. The geometry is defined using the following steps.
\begin{itemize}
\item \textbf{Spaxel Location:} The spectral traces, or spaxels, are defined using the dome flats.  A series of exposures of the dome flat patch, illuminated by a halogen lamp, is taken and combined to remove cosmic rays and improve the S/N of the image.  \texttt{SExtrator} is used to identify each spaxel segment in the image.  These segments each have a unique segment ID and are used to derive the geometric properties of each spaxel.  Each spaxel is fit with a second order polynomial to derive the center of each spaxel as a function of CCD X and Y pixel position.  The width of the spaxel is retained as well.  Since the blue illumination from the halogen lamp is low, the fits are extrapolated into the blue.  Using a very low order polynomial ensures that this will not produce spurious results.  The master dome flat image with spaxel fits overlaid is shown in Figure \ref{fig:domesegments}.
\begin{figure}[h]
\begin{center}
\centering
\includegraphics[width=0.48\textwidth]{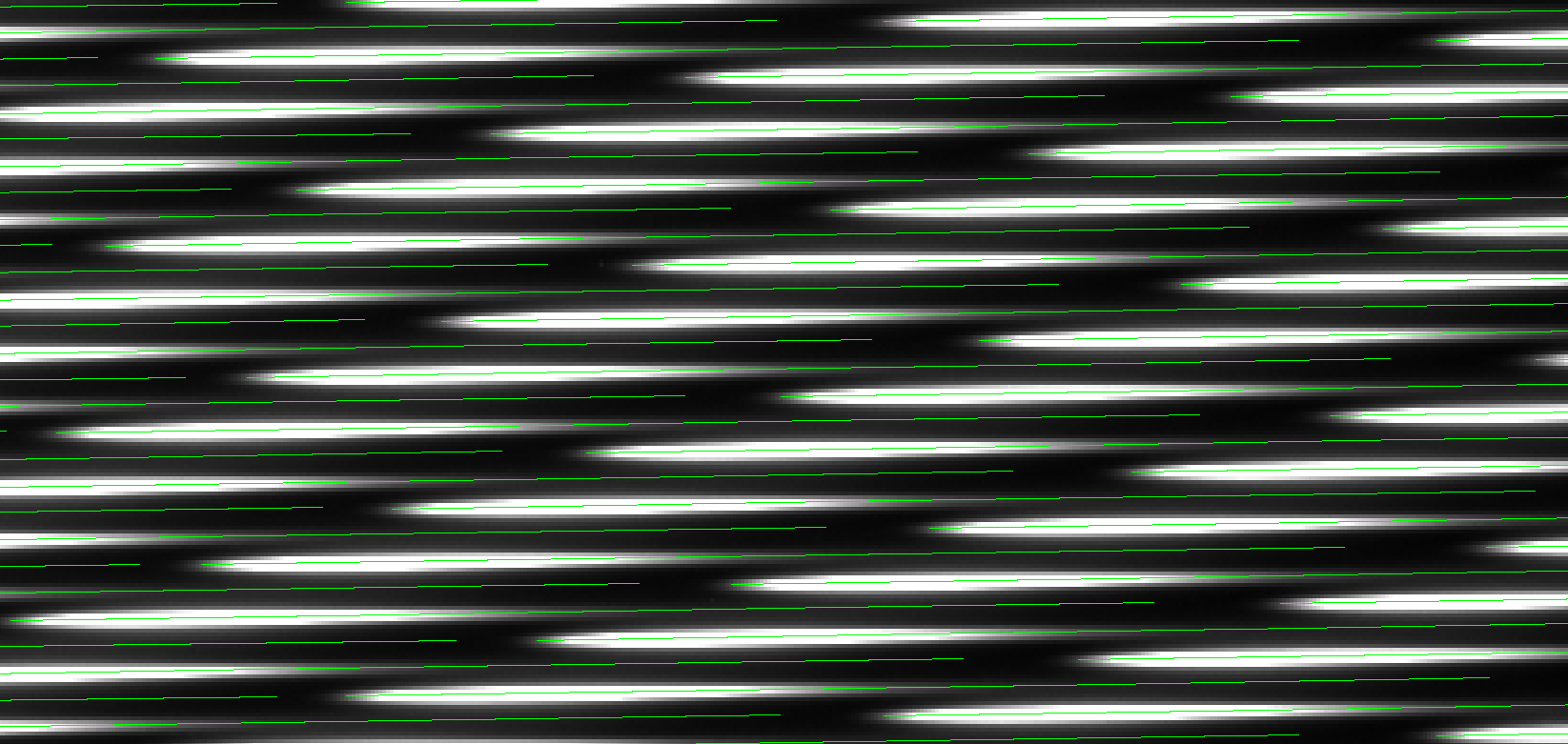}
\end{center}
\caption{ Zoom of dome flat master image with trace fits over-plotted in green.}
\label{fig:domesegments}
\end{figure}

\item \textbf{Rough Wavelengths:} Seven Hg arc lamp images are combined to form a master image.  The brightest Hg lines are then identified in each spaxel trace, yielding a preliminary wavelength solution for each trace.
\item \textbf{Fine Wavelengths}: Seven each of Xe and Cd arc images are combined to yield master Xe and Cd images that are analyzed to produce a refined wavelength solution for each trace.  The median RMS for the wavelength fit residuals of all spaxels is typically 0.3 nm.
\item \textbf{Spatial Projection:} This fine solution is then de-projected using the known geometry of the hexagonal lenslets to derive an X,Y position for each trace, relative to the spaxel closest to the center of the array, at a specific wavelength in the field of view of the IFU.  These coordinates are recorded in pixels relative to the detector frame and in arcseconds relative to the center of the IFU.
\end{itemize}

\subsubsection{Object Extraction}
Once the geometric mapping of the spatial and spectral dimensions are defined, science observations can be reduced and spectrally extracted using the following steps.

\begin{itemize}
\item \textbf{Flexure Correction:} As the telescope is moved around the sky, the  flexure in the spectrograph can cause the spectrum to shift on the detector with small offsets. In order to derive this value, science images are first registered with the spatial and spectral geometry solution for the night.  This is done in the Y-pixel direction by comparing spaxel locations in the master dome flat with those in the science images as delineated by the sky spectra to derive a master flexure offset in the Y-direction.  This shift has a typical intra-night variation of $\simeq$0.5\,pix.  For the X, or spectral direction, flexure is measured using one or both of two bright, relatively isolated sky lines at 557.7 nm (OI) and 589.6 nm (Na).  This offset is typically less than 2 nm and the two lines (when both are measurable) typically agree to within the errors in the wavelength solution (0.3 nm).  All subsequent extraction operations use these flexure correction values to extract the spectrum from a given science image.

\item \textbf{Background Subtraction:}  An iterative background subtraction method is applied to remove scattered light from the IFU images.  First, the flexure corrected spaxel locations are used to mask IFU image pixels that have been directly illuminated from the IFU.  The unmasked region is assumed to contain only scattered light and is convolved with a 2D smoothing Gaussian kernel of 17 pixels width with successive surface fits for five iterations until a good map of scattered light is obtained.  This is written out and subtracted from the science image.

\item \textbf{Sky Subtraction}:  For fainter objects ($R>16.5$\,mag), a pair of images is obtained with a small ($\approx7$\arcsec) pointing offset such that the object in one image falls on a background region in the other.  The two images are then subtracted from one another to improve the target's contrast against the sky.

\item \textbf{Spaxel Extraction:}  All the spaxels for a given image are extracted by using the traces as defined by the master dome and corrected by the flexure values.  A raw spectrum for each spaxel is built up by moving along the spaxel trace one pixel at a time in the X or spectral direction and performing a weighted sum of the image spaxels in the Y direction extending above and below the trace center by three pixels (for a total of seven pixels).  The weighting is calculated from a normalized Gaussian with same $\sigma$ as the trace in the master dome flat image.  This weighting minimizes the contamination from neighboring spaxels.  The flat field value is also divided out at this time.  The spaxels for the variance image are extracted with the identical method so the SNR can be calculated for each extracted spectrum.

\item \textbf{Object/Sky-Host Location:} Light for a given object falls on a subset of all the spaxels.  The object spaxels must be defined along with a set of sky/host light spaxels.  These spaxels are then used as a statistical ensemble to derive the best sky/host-subtracted spectrum.  Prior to object selection, a pseudo-image is formed by collapsing all the spaxels in wavelength from 650 to 700 nm, near the peak response of the IFU system.  For bright ($R<16.5$\,mag) objects, an automatic object detection algorithm is then applied to the pseudo-image and fit with a 2D Gaussian to derive the extraction aperture.  The sky spaxels are then defined based on the object extraction aperture.  If the object is a standard star, which is defined to be a bright, isolated star, then all non-object spaxels are used to derive the sky spectrum, after applying a 3-sigma rejection criterion.  For all other objects, the sky annulus is defined by extending the object aperture by 3\arcsec and excluding the object spaxels as defined by the object aperture.  This is applied even to A and B image pairs despite of the sky subtraction already performed due to the need to also subtract host galaxy light.

\item \textbf{Interactive Aperture Placement:}  Since the offsets for the A/B pair images are adjusted for each individual object, and due to the small field of view of the IFU, which precludes an astrometric solution, these apertures are currently placed interactively.  A graphical user interface (GUI) displays the pseudo-image with an aperture that tracks the mouse movement.  Once the aperture is correctly placed, based on finder charts of the object, a mouse click registers the object location.  The GUI also provides key-strokes to expand or shrink the aperture.

Figure \ref{fig:ifuhost} shows the capabilities of the SEDM to extract both object and host spectra.  All spaxels within the field of view are eligible for extraction.  At the top left, a Type Ia SN, iPTF15drk, is shown in the SEDM pseudo-image with the A and B apertures indicated.  The resulting spectrum is shown in the top right plot.  The bottom left plot shows the apertures for the host galaxy and the bottom right shows the resulting spectrum.  This capability is useful for checking the redshift derived from the SN and also for checking the host subtraction.

\begin{figure}[h]
\centering
\includegraphics[width=\columnwidth]{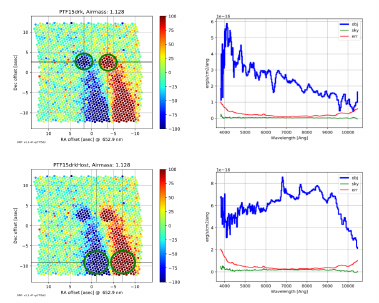}
\caption{Extraction of SN and Host spectra for iPTF15drk, a Type Ia SN.  SN is at the top and host is at the bottom.  The color scale for the images on the left is arbitrary relative intensity, with blue values being negative and red values positive.}
\label{fig:ifuhost}
\end{figure}

\item \textbf{Spectrum Extraction:} If the object is a standard star, then the Gaussian fit is extended to include 99\% of the light and each spaxel is used to calculate the mean spectrum.  If the object is not a standard, then the flux for each spaxel is summed between 500 and 900 nm and this total flux is used to rank order the spaxels.  The brightest 70\% of the spaxels are retained for determining the final spectrum.  This prevents the faintest spaxels from adding noise to the final spectrum.  The mean spectrum is then calculated for both the object and sky/host spaxels.  The sky/host spectrum is then subtracted from the object spectrum and the object spectrum is scaled by the number of spaxels used.  The error spectrum is calculated from the variance image spaxels corresponding to the object and sky spaxels used in the object image.
\end{itemize}

\subsection{Calibration}

Spectrophotometric standards are observed and extracted as defined in the previous subsection.  The reference flux for the standard is re-sampled onto the SEDM IFU wavelength scale, extinction corrected,  and compared with the instrumental spectrum of the standard to derive an inverse sensitivity curve.  This can then be used to convert any instrumental spectrum into a flux-calibrated spectrum.  Anywhere from one to five spectrophotometric stars are observed during the night, and one generally taken at the beginning of the night during late twilight.  The ensemble of calibrations are combined to form the mean calibration after each standard star is observed.


\section{Performance and first results} \label{sec:results}
\subsection{Rainbow Camera} \label{sec:rc_performance}

\begin{figure}[h]
\begin{center}
\centering
\includegraphics[width=0.5\textwidth]{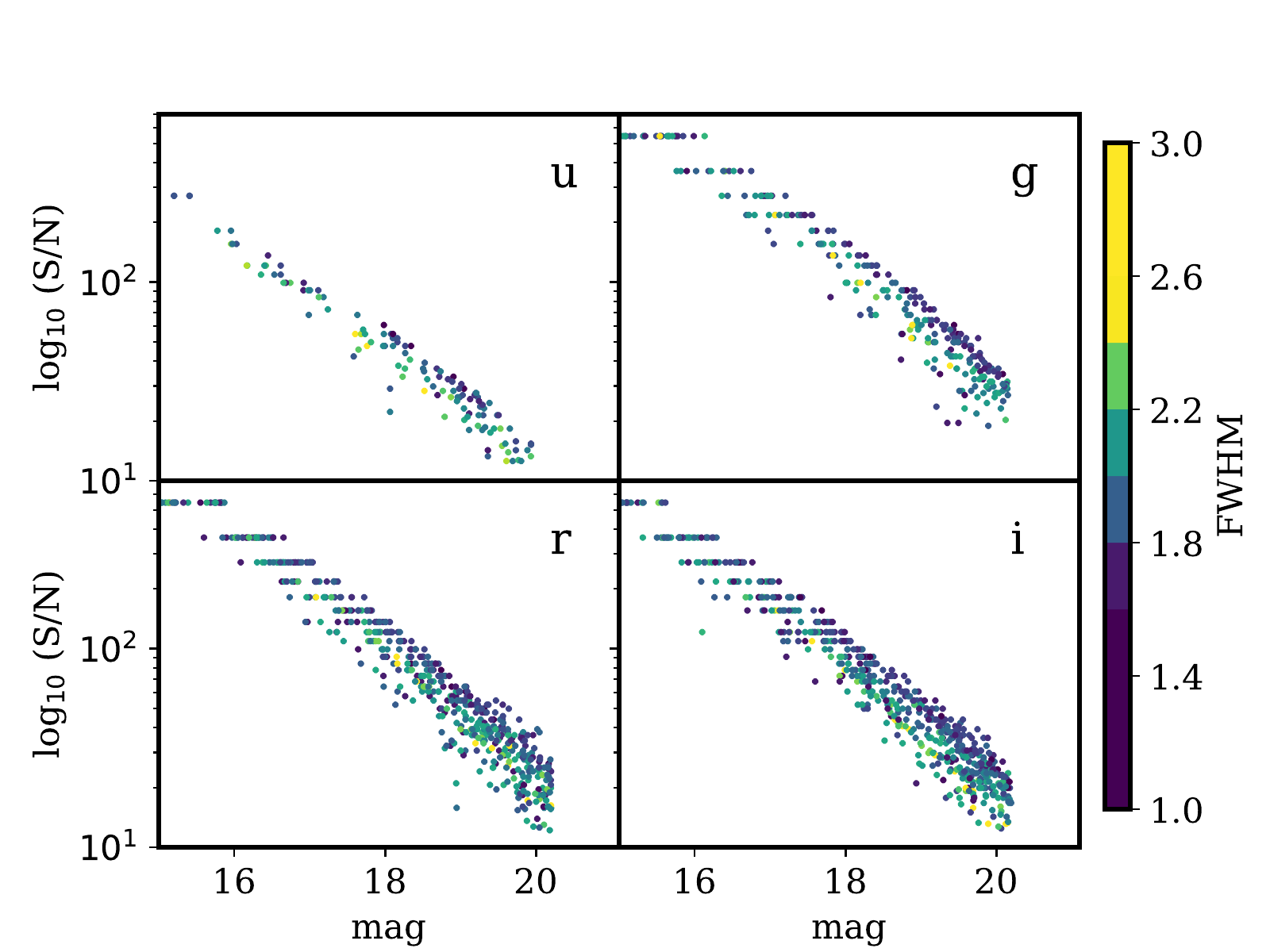}
\end{center}
\caption{ SNR for each filter for a 180\,s exposure. The color indicates the FWHM on the exposure. Only observations with airmass $<$3 have been included into the plot.}
\label{fig:signal2noise}
\end{figure}

Figure \ref{fig:signal2noise} shows the SNR for objects with magnitudes 13-21 in different filters. The average S/N for each filter and magnitude bin is shown in Table \ref{table:s2n}.

\begin{deluxetable}{lcccc} 
\tablecolumns{2} 
\tablecaption{Average S/N for stars within each magnitude bin. We include only isolated point-like sources. The results are computed for an exposure time of 180\,s. \label{table:s2n}} 
\tablehead{ 
\colhead{ filter} & \colhead{17 $-$ 18} & \colhead{18 $-$ 19} & \colhead{19 $-$ 20} & \colhead{20 $-$ 20.5}  \\
\colhead{ } & \colhead{[mag]} & \colhead{[mag]} & \colhead{[mag]} & \colhead{[mag]}  } 
\startdata 
u$\prime$ & 55 & 30 & 20 & 12 \\
g$\prime$ & 180 & 85 & 40 & 30 \\
r$\prime$ & 135 & 70 & 30 & 23 \\
i$\prime$ & 120 & 60 & 30 & 20 \\
\enddata 
\end{deluxetable}

\subsection{Integral Field Unit} \label{sec:ifu_performance}

\begin{figure}
\centering
\includegraphics[width=0.8\columnwidth,angle =270 ]{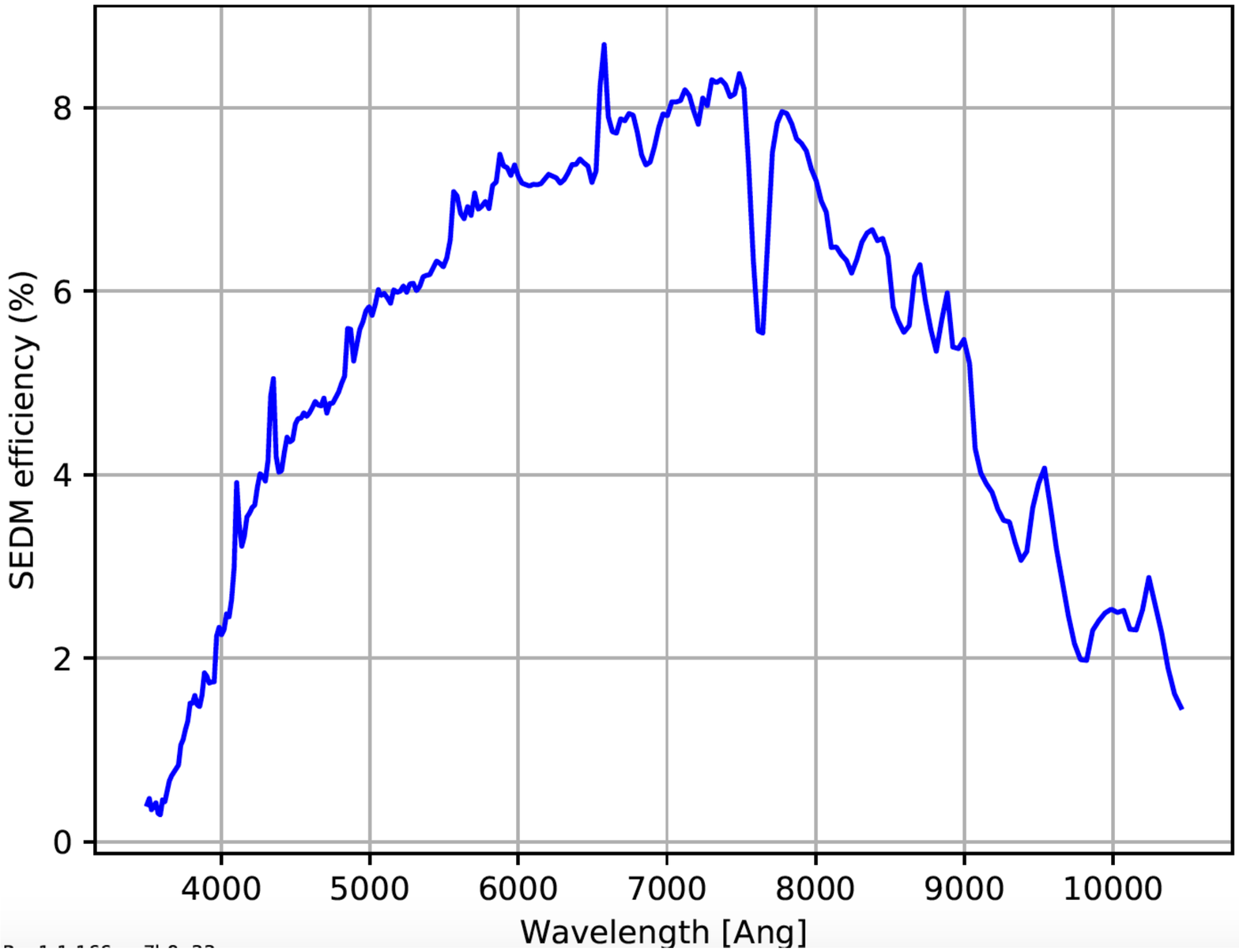}
\caption{IFU efficiency in percent as a function of wavelength in \AA\  derived from a 300s exposure of the standard star HZ2.  We assume an area of 18,000 cm$^2$ for the P60 and an overall reflectivity of 82\%.}
\label{fig:ifueff}
\end{figure}

The IFU was designed to cover a wavelength range from 350 to 1000 nm.  By comparing standard star observations with their calibrated fluxes, we find that the IFU total system response (telescope excluded) peaks at $\sim 8$\% between 700 and 800\,nm (see Figure \ref{fig:ifueff}).  The blue response is lower than expected from the optical models, and its causes are discussed int eh next Section.  As a result, we typically do not use wavelengths below 400\,nm for object classification. 

The average S/N for the IFU is shown in Table \ref{table:ifus2n}. The S/N depends not only on the system throughput but also on any component of scattered light we have in the system. Because of this scatter, we require to use the A/B method for sky subtraction, which currently reduces our sensitivity by a factor of $\approx\sqrt{2}$.

\begin{figure}
\centering
\includegraphics[width=0.9\columnwidth]{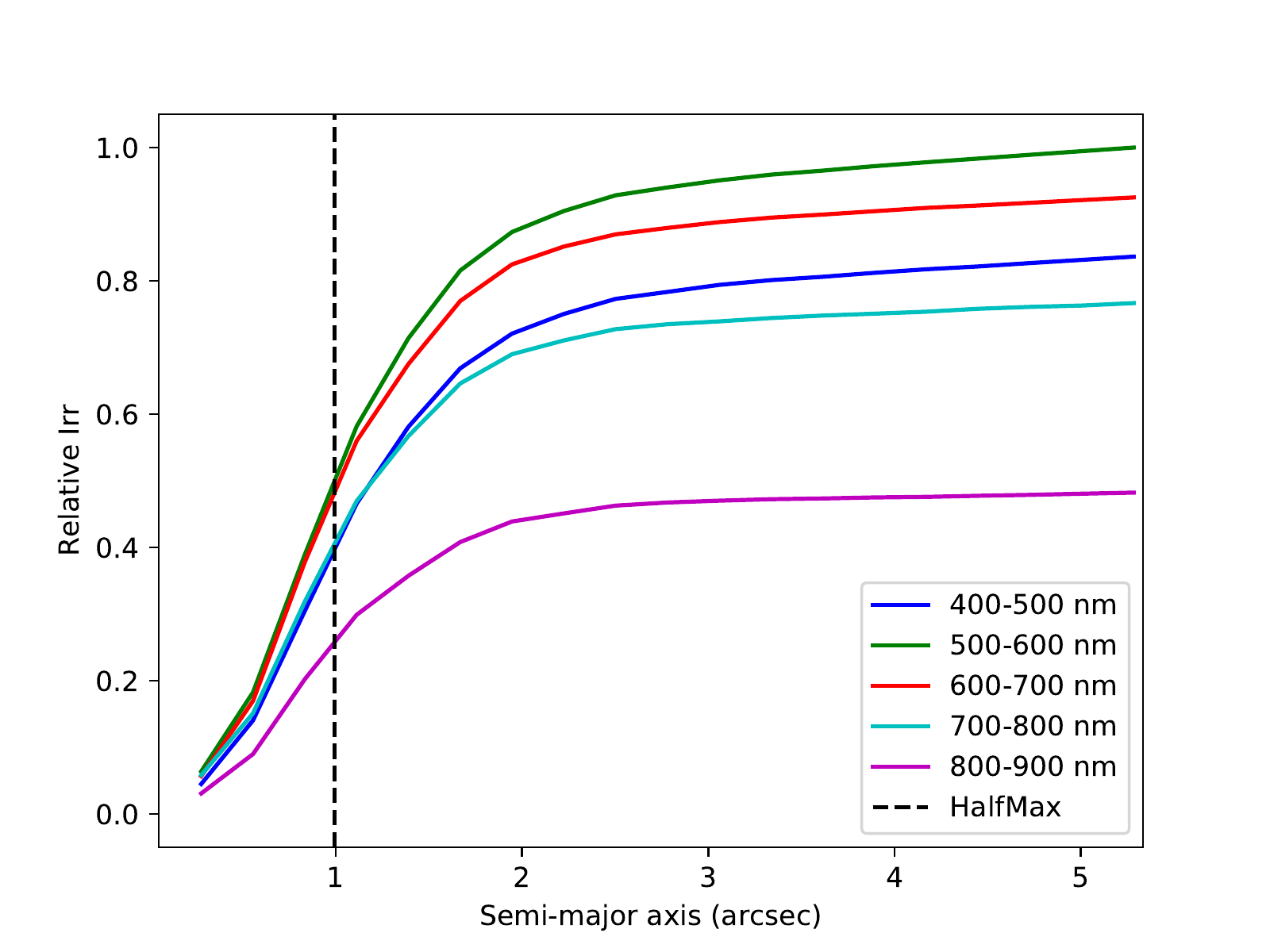}
\caption{Curve of growth for IFU observation of Feige 34 for five wavelength bins.  The half-light radius is indicated by the vertical dashed line and is equivalent to a seeing FWHM of 2 \arcsec. }
\label{fig:ifucog}
\end{figure}

The lenslet separation projected on the sky is 0.64\arcsec in the RA direction and 0.56\arcsec in the Dec direction.  For the median seeing on P60 of 1.68\arcsec, we expect the point-spread-function (PSF) to be oversampled, which can also lead to an additional reduction in S/N.  Figure \ref{fig:ifucog} shows the curve of growth for a standard star normalized to the $500-600$ nm bandpass.  Half of the light for this bandpass is contained within an aperture with a semi-major axis of 1$"$, which is equivalent to a FWHM of 2\arcsec.

We calculate the SNR of the extracted SEDM spectra in the following manner.  During reduction, we create a variance image which includes the Poisson noise from the signal and the read noise per pixel.  This is calculated by simply adding the square of the read-out noise ($5\times5$ = 25e$^{-}\,^{2}$ to the input bias-subtracted image.  When the object and sky spaxels are identified prior to extraction, these same spaxels in the variance image are used to calculate the extracted variance spectrum.  The object variance and the sky variance are added together so we account for both object and sky sources of uncertainty (unless we turn off the sky subtraction, in which case we just use the object variance).  The noise in the spectrum is then simply the square root of the variance spectrum. This estimation does not include systematic errors, such as the ones cause by the scattered light.

As stated in \S\ref{sec:intro}, our design goal for the SEDM was to achieve a SNR of $\sim$5 for an object with an $r$-band magnitude of 20.5 in an exposure of 3600\,s.  Table \ref{table:ifus2n} gives the average SNR for 43 extracted spectra for $r$-band magnitudes from 16 to 20 in each of three wavelength bins.  This shows that we are consistently achieving the target SNR for objects with $r$ magnitudes of 19.5, however, it is unlikely that we are achieving this goal for objects with $r$-band magnitudes fainter than 20.5.  The 43 objects used for this figure span a range of types: from bright, stellar, non-variable sources to fainter, host-involved transients.  The scatter in the plot and the variation for a given waveband are due to the variation in the input spectra and the variation in host involvement, which also affects SNR.

\begin{deluxetable}{lcccc} 
\tabletypesize{\footnotesize} 
\tablecolumns{5} 
\tablecaption{Average S/N for objects within each magnitude bin. The results are computed for an exposure time of 3600\,s. \label{table:ifus2n}} 
\tablehead{ 
\colhead{ Wl. Range } & \colhead{16 $-$ 17} & \colhead{17 $-$ 18} & \colhead{18 $-$ 19} & \colhead{19 $-$ 20}  \\
\colhead{ [nm] } & \colhead{[mag]} & \colhead{[mag]} & \colhead{[mag]} & \colhead{[mag]}  } 
\startdata 
400-500 & 11.3 & 10.2 & 7.2 & 3.4 \\
500-800 & 20.3 & 18.0 & 10.1 & 5.5 \\
800-900 & 18.6 & 15.9 & 5.8 & 3.1 \\
\enddata 
\end{deluxetable}

\subsection{Improving SEDM IFU optical transmission}

The optical modeling of the throughput for IFU predicted an efficiency  $>$ 40\% from 450\,nm $-$ 750\,nm. However, the measured efficiency on-sky appears to be only ~5\% at 450\,nm and ~8\% at 750\,nm (see Figure \ref{fig:ifueff}), with an even larger deficiency at wavelengths shorter than 400\,nm.  An investigation of this situation was conducted in summer 2017 at the Caltech Optical Observatory laboratories. The procedure involved the total disassembly of the optical system and the evaluation of the spectral transmission values for each component.  The camera and optical collimator assemblies were tested as units, so as to not disturb their internal alignments.  

This exercise determined that the integral-field hexagonal lenslet array contributed the plurality of all optical losses.  This array has a corner-to-corner diameter of 592$\mu$m (0.74\,\arcsec) and focal length of 2660\,$\mu$m, resulting in a fast focal ratio of f/4.5.  Closer examination of this array showed that there was an approximately 62\,$mu$m (radial) width \textit{deadband} on either side of the lenslet boundaries. Optical light falling on this ``gap'' is not cleanly directed into desired micro-pupils, but rather is extensively scattered.  The final measured median image size at the lenslet array is 2.2\,\arcsec\ FWHM, implying that the \textit{deadband} scatters ~20\% of the total area of the image in an unpredictable manner.  Such scatter removes signal from the micro-pupils, but also widely disperses noise photons into the spectrograph focal plane.  A review of the original vendor procurement specification called out a maximum fill-factor of 95\%, suggesting that the usable areal fraction of each lenslet was fabricated on a best-effort basis, and only 80\% effective fill factor was obtained.

In response to these investigation, we intend to seek a new alternative lenslet vendor capable to delivering smaller \textit{deadband}.  While the original hexagonal lenslet design choice was made in an overt attempt to ease the optical fabrication requirement for vendors, we believe greater experience within the community with high-fill-factor square lenslet arrays now warrants investigation of such a design change. At the time of this writing, the vendor \texttt{JENOPTIK}\footnote{\url{www.jenoptik.com}} has quoted a minimum 98\% fill-factor square f/4.5 lenslet for SEDM. Currently, we are seeking independent measurement verification of this performance to evaluate its suitability.

\section{Examples and early science} \label{sec:examples}

As noted in \S\ref{sec:intro}, routine operations began in late April 2016. The SEDM has been in continuous use until end of February 2017, following which the instrument was taken off for improvements. Over this period of 10 months
the SEDM has obtained spectra of 1660 objects of which 465 corresponded to iPTF transient objects. The usage of the instrument has allowed us to increase the number of classifications in 25-40\% regarding previous years.

 Currently, the dozen or so spectroscopic targets for SEDM observations are specified on a nightly basis. However, a major strength of the SEDM arises is its low latency $\Delta t$, defined as the time between a request for observations and spectral classification. In a few months we intend to implement a target-of-opportunity mode in which requests for SEDM observations can be made and observations undertaken within minutes (e.g.\ GRB afterglow). 
 
\begin{figure}[ht]
\centering
\includegraphics[width=\columnwidth]{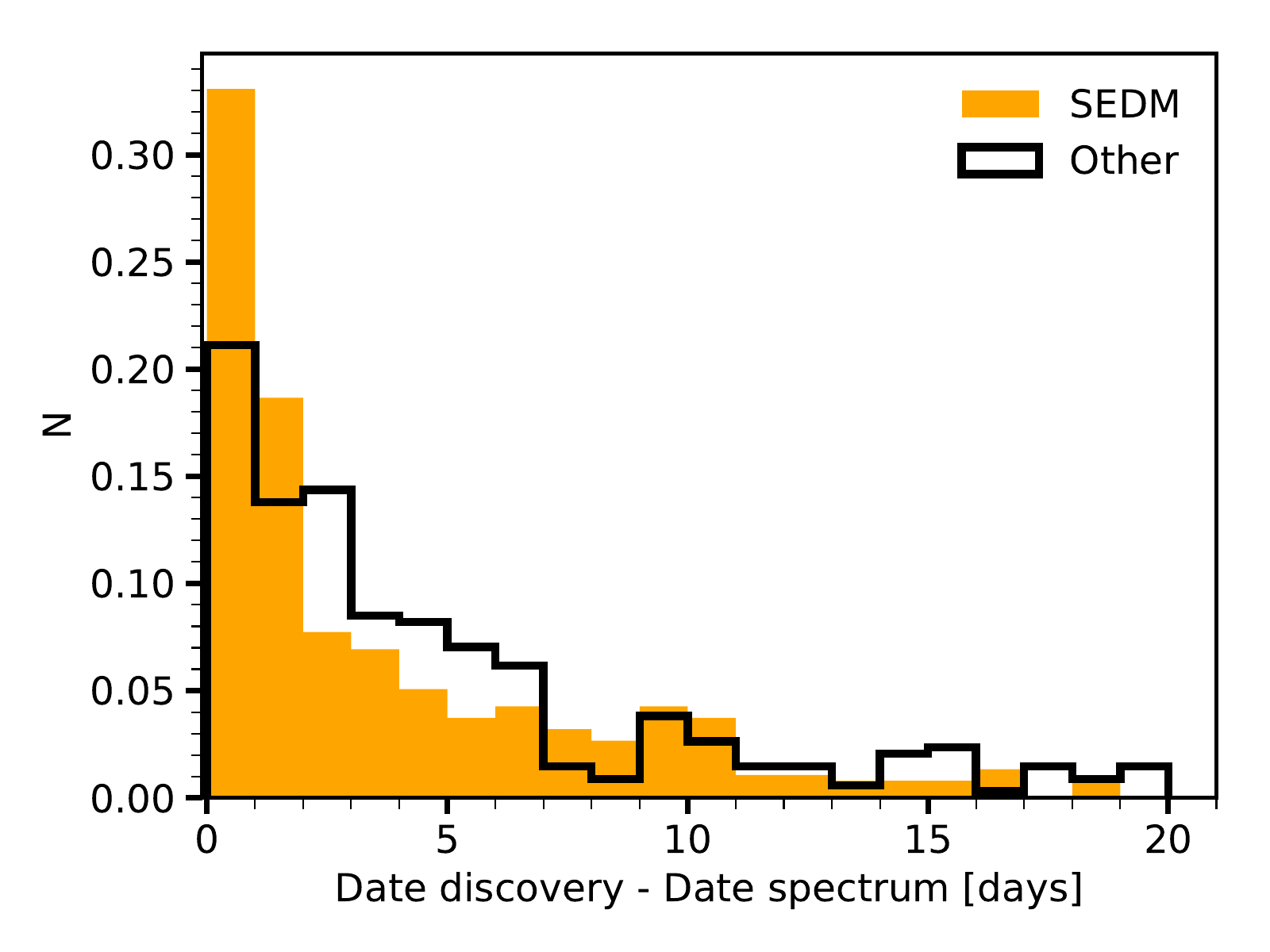}
\caption{Normalized histogram of the time elapsed between the date of discovery for a transient, and the date of the first spectrum. Orange bars show the distribution of time for transients with SEDM spectra and the black line for other telescope facilities. The data for the latter were obtained from the PTF transient database.} \label{fig:delay}
\end{figure}

The latency of the observations over the last two months is 
shown in Figure \ref{fig:delay}. For all transients with an SEDM spectrum, around a third of them were observed within a day of their discovery, half of them in less than 2.5 days and a two thirds within the first 5 days of discovery. Some of the delays in obtaining a spectrum are due to inclement weather conditions while others are related to human selection criteria or the shared usage of SEDM with other research groups. Overall, if we compare $\Delta t$ for other transients observed with spectrographs on larger telescopes available to the PTF group (i.e. Palomar Hale telescope, Keck, DCT, Gemini, NOT or APO), only 20\% were classified within a day of their discovery, and the median delay time raises to 4 days.

\begin{figure*}[ht]
\centering
\includegraphics[width=2\columnwidth]{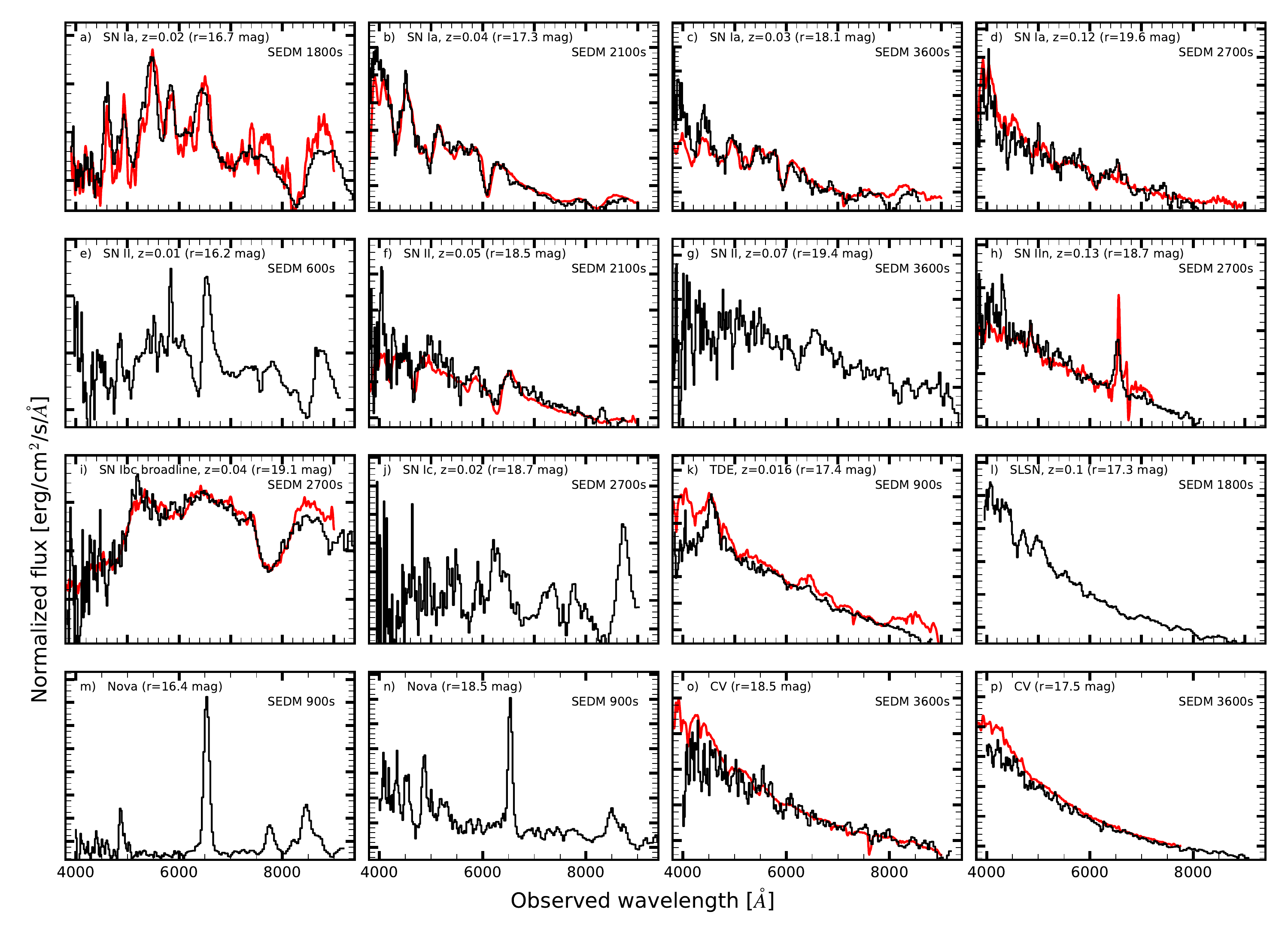}
\caption{Sample of SEDM spectra for different transient types (black thick lines). Whenever available, long slit spectroscopy taken at another telescope at a similar epoch is shown as a red thin line.}
\label{fig:specexample_transients}
\end{figure*}

Among the classified transients, as expected, the most common types were SN type Ia, followed by SN type II, AGN, stellar sources and cataclysmic variables (CVs). An example of these spectra is shown in Figure \ref{fig:specexample_transients} and the spectral log is shown in Table \ref{tab:speclog}. Whenever available, the Figure also displays a spectrum taken at similar epoch with another instrument for comparison. The sample includes SN type Ia with different magnitude ranges, SN type II and SN Ic. It also contains novae, CVs, and a couple of less frequent transients, such as a super-luminous supernova (SLSN) and a TDE. 

The most notable discoveries of the SEDM are: one of closest optical tidal disruption events \citep[iPTF16fnl;][]{Blagorodnova2017ApJ}, an unusual subclass of type Ia supernova 02cx-like  \citep{Miller2017arXiv}, a rapidly turning-on quasar \citep[iPTF16bco;][]{Gezari2017ApJ} and a lensed SN type Ia \cite[iPTF16geu;][]{Goobar2017Sci}. 

\begin{figure*}[ht]
\centering
\includegraphics[width=2\columnwidth]{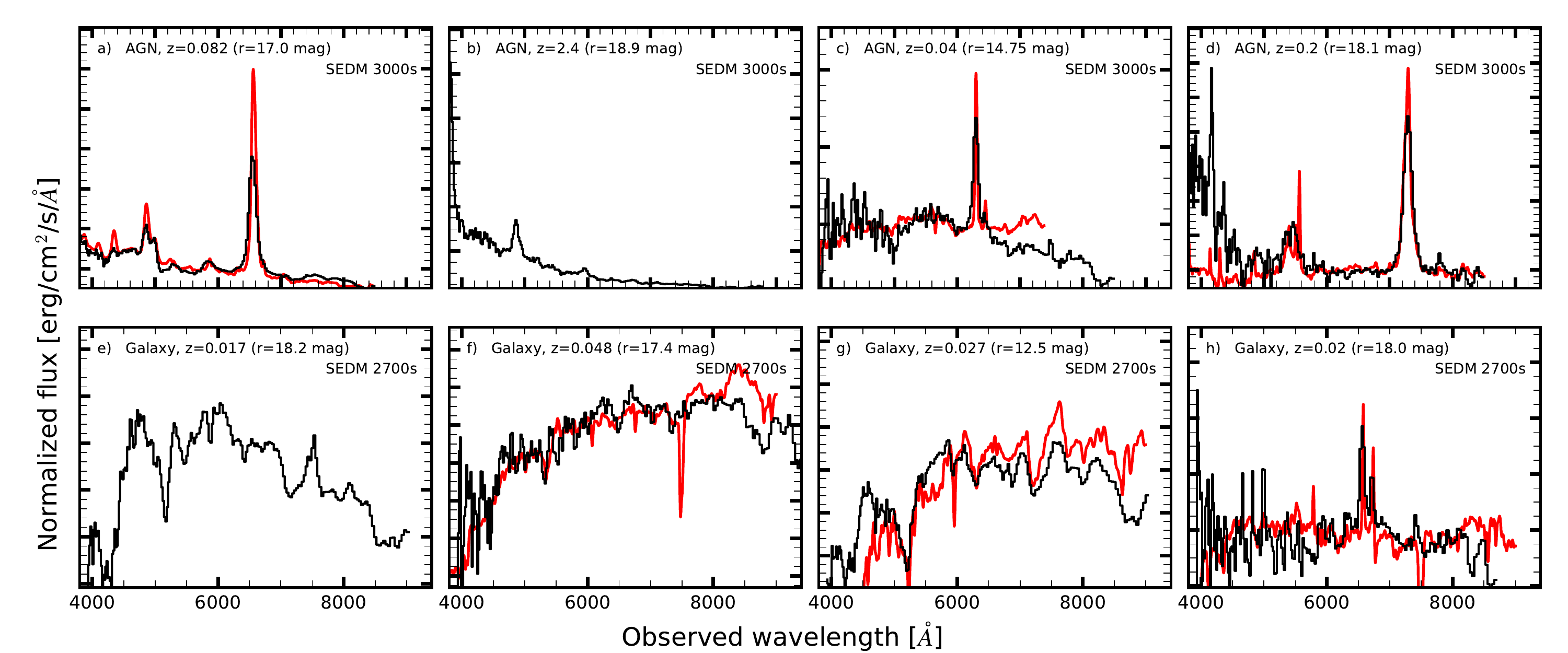}
\caption{Sample of SEDM spectra (black line) for different AGN and galaxies. Whenever available, long slit spectroscopy taken at a similar epoch is shown as a red thin line.}
\label{fig:specexample_galaxies}
\end{figure*}

The SEDM can also be used for confirmation of redshifts of nearby ($\lesssim 200\,$Mpc) star forming galaxies. The typical precision (1$\sigma$) for the recession velocity inferred from the SEDM spectra is $1500\,{\rm km\,s^{-1}}$ and the accuracy is $\sim$250\kms.
Figure \ref{fig:specexample_galaxies} shows several examples of AGN and spectra of galaxies at different magnitudes. The spectra are adequate enough to allow us to distinguish between passive and active galaxies.

\begin{figure*}[ht]
\centering
\includegraphics[width=2\columnwidth]{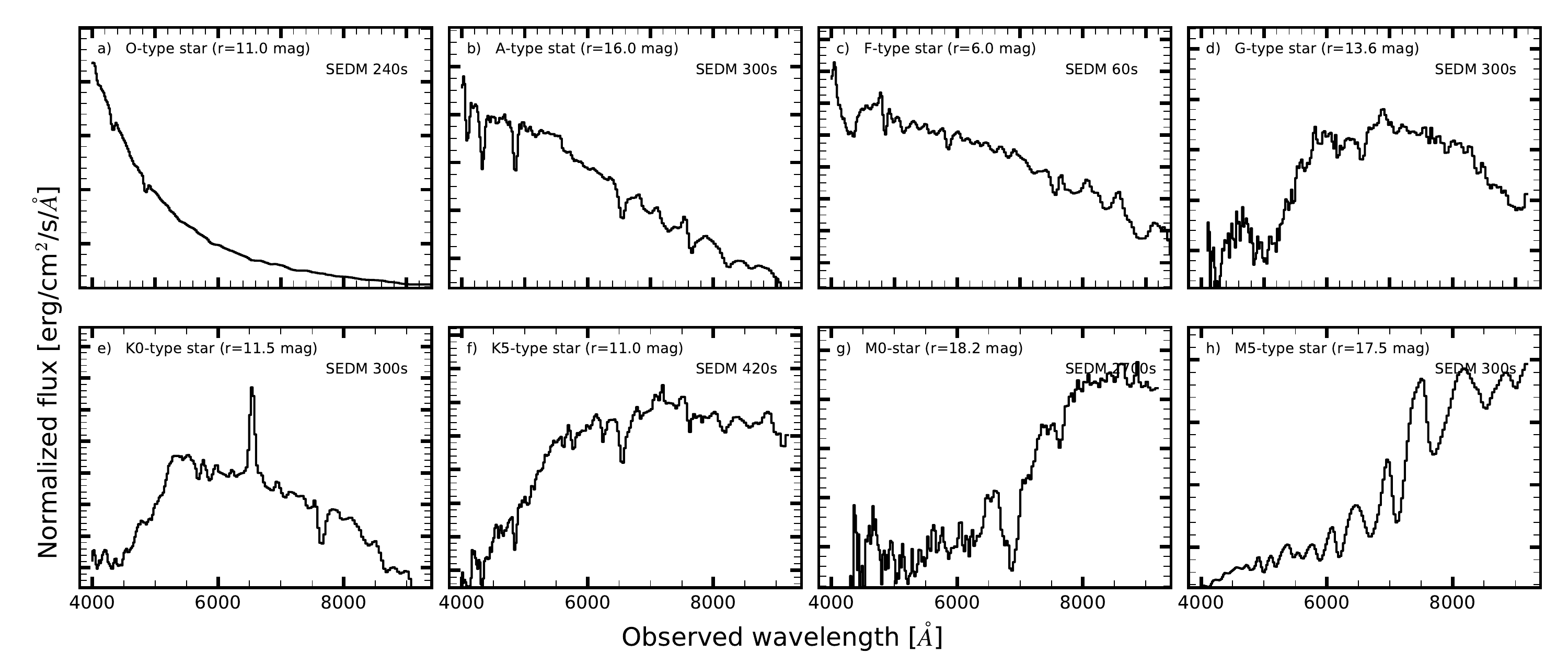}
\caption{Sample of SEDM spectra for different stellar objects (black thick lines). Generally, the stellar sources have magnitudes between 11 and\,18 mag. Panel e) shows an example of an emission line star in the open cluster NGC 743.}
\label{fig:specexample_stars}
\end{figure*}

\begin{figure}[h]
\begin{center}
\centering
\includegraphics[width=0.5\textwidth]{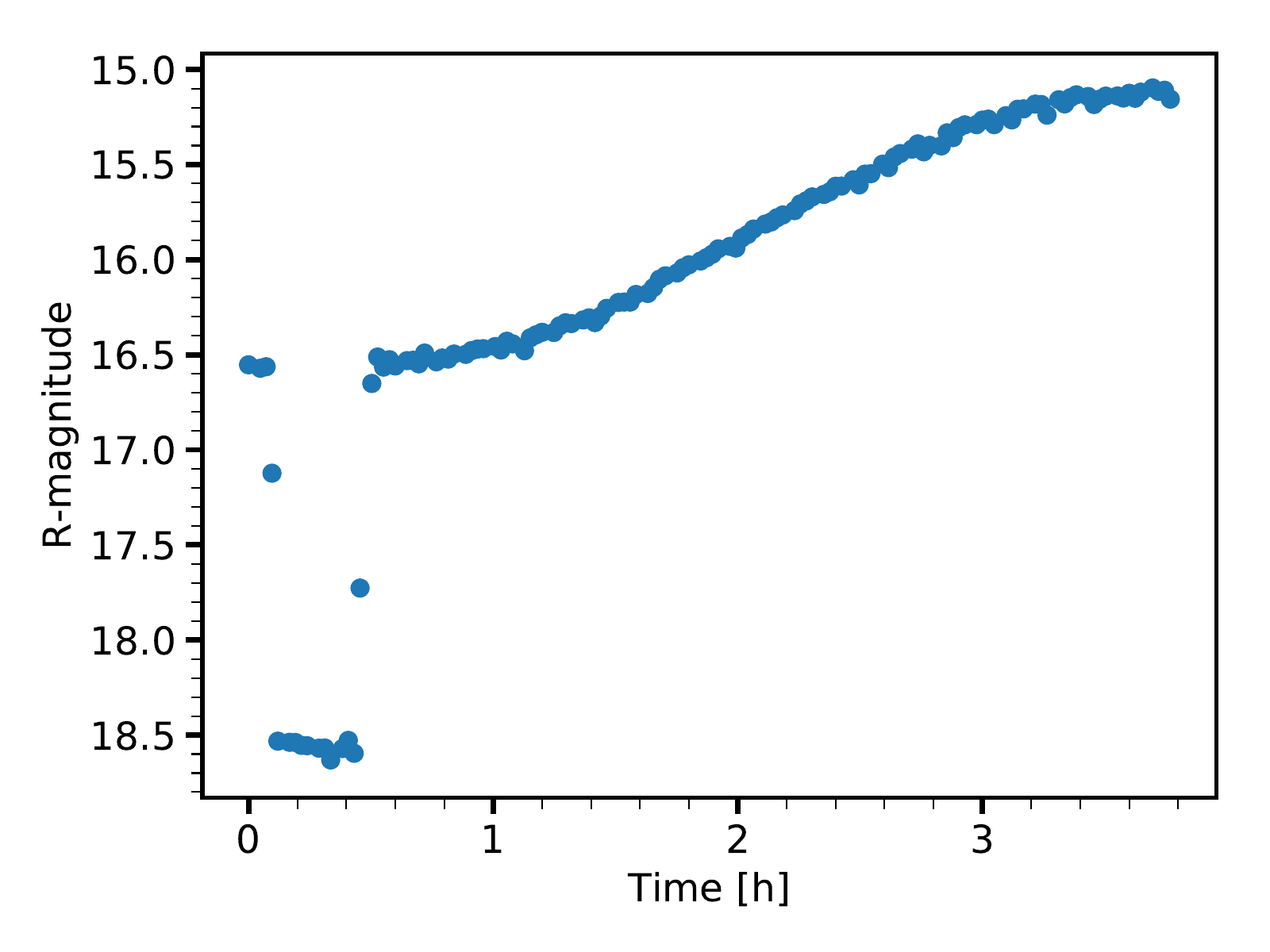}
\end{center}
\caption{Lightcurve of an eclipsing binary system. The data shown was obtained with the SEDM aperture photometry pipeline. The exposure time was fixed to 60\,s. The error bars are smaller than the size of the markers.}
\label{fig:variable_lightcurve}
\end{figure}

The SEDM has been used to study stars in two different ways: photometry and spectroscopy.
Examples for the former include detailed  characterization of eclipses in close binary systems (T. Kupfer et. al. in prep) such as the one shown in Figure \ref{fig:variable_lightcurve}, multi-band follow-up of binaries showing reflection effects from a hot compact companion and complementary observations for determination of periods for variable stars. An example of spectroscopic observations of stellar sources is shown in the collage in Figure\,\ref{fig:specexample_stars}.
The surprising success of the SEDM in determining effective temperature (via model fits to the data) have motivated us to embark on a project to quantitatively determine the precision of the SEDM in measuring the effective temperature as well as the degree of reliability of spectral classification.  
Additionally, the SEDM has also enabled a search for emission lines sources. Two new Be stars in the open cluster NGC\,6830 were identified via their emission lines in H$\alpha$ and absorption lines of H$\beta$, H$\gamma$, HeI\,4471 and MgII~4481 \citep{Yu2016AJ}.

\begin{table*}
\centering
\begin{small}
\caption{Spectroscopic log for Figures \ref{fig:specexample_transients}, \ref{fig:specexample_galaxies} and \ref{fig:specexample_stars}.}

\begin{tabular}{ccccccc}
\hline
Panel & Classification  & Redshift & Magnitude & $t_{SEDM}$ & Telescope & $t_{tel}$   \\ 
    	&  				& 			&	(mag) &   (s) &+Instrument & (s)  \\ \hline
        \multicolumn{7}{c}{Figure \ref{fig:specexample_transients}} \\ \hline
a &    SN Ia &  0.02  & r=16.7  & 1800 & FTN & 2700 \\
b &    SN Ia &  0.04  & r=17.3  & 2100 & P200 & 300 \\
c &    SN Ia &  0.03  & r=18.1  & 3600 & P200 & 300 \\
d &    SN Ia &  0.12  & r=19.6  & 2700 & P200 & 600 \\
e &    SN II &  0.01  & r=16.2  & 600 & -- & -- \\
f &    SN II &  0.05  & r=18.5  & 2100 & P200 & 300 \\
g &    SN II &  0.07  & r=19.4  & 3600 & -- & -- \\
h &    SN IIn &  0.13  & r=18.7  & 2700 & DCT & 900 \\
i &    SN Ibc BL &  0.04  & r=19.1  & 2700 & P200 & 600 \\
j &    SN Ic &  0.02  & r=18.7  & 2700 & -- & -- \\
k &    TDE &  0.016  & r=17.4  & 900 & P200 & 900 \\
l &    SLSN &  0.1  & r=17.3  & 1800 & -- & -- \\
m &    Nova &0 & r=16.4 & 900 & -- & -- \\
n &    Nova &0 &r=18.5  & 900 & -- & -- \\
o &    CV & 0 &r=18.5  &  3600 & P200 & 1500 \\
p &    CV & 0 &r=17.5  & 3600 & DCT & 1200 \\ \hline

        \multicolumn{7}{c}{Figure \ref{fig:specexample_galaxies}} \\ \hline
a &    AGN &  0.082  & r=17.0  & 3000 & Keck1 & 300 \\
b &    AGN &  2.4  & r=18.9  & 3000 & -- & -- \\
c &    AGN &  0.04  & r=14.75  & 3000 & DCT & 600 \\
d &    AGN &  0.2  & r=18.1  & 3000 & SDSS+2.5M & 6000 \\
e &    Galaxy &  0.017  & r=18.2  & 2700 & -- & -- \\
f &    Galaxy &  0.048  & r=17.4  & 2700 & P200 & 1200 \\
g &    Galaxy &  0.027  & r=12.5  & 2700 & SDSS+2.5M & 3600 \\
h &    Galaxy &  0.02  & r=18.0  & 2700 & P200 & 1200 \\ \hline

        \multicolumn{7}{c}{Figure \ref{fig:specexample_stars}} \\ \hline
a &    O-type & 0 & r=11.0 & 240 & -- & -- \\
b &    A-type  & 0 &r=16.0  & 300 & -- & -- \\
c &    F-type & 0 &r=6.0  & 60 & -- & -- \\
d &    G-type & 0 &r=13.6  & 300 & -- & -- \\
e &    K0-type & 0 &r=11.5  & 300 & -- & -- \\
f &    K5-type & 0 &r=11.0  & 420 & -- & -- \\
g &    M0-type & 0 &r=18.2  & 2700 & -- & -- \\
h &    M5-type & 0 &r=17.5  & 300 & -- & -- \\

\hline
\end{tabular}
\label{tab:speclog}
\end{small}
\end{table*}

\section{Discussion and conclusions} \label{sec:discussion}

We have presented an overview of the instrumental design and operations of SEDM, a dedicated low resolution IFU spectrograph on the Palomar 60-inch telescope. The instrument has been used for spectroscopic classification and photometric follow-up of iPTF transients discovered during 2016 and early 2017. Overall, the SEDM produces consistently classifiable spectra for targets brighter than $R = 19.5$ (in 1\,h exposure). Although iPTF limiting magnitude is about 1\,mag deeper than that, discoveries made during PTF/iPTF era show that around 1/3 of all transients were detected with magnitudes $<$19.5 in either $g$ or $R$-band. Around 40\% of all transients were brighter than 19.5\,mag at peak, and therefore suitable for classification with SEDM. With an average rate of 10 spectra per night, the SEDM has proven to be a useful tool for the fast initial classification of candidates and photometric follow-up. 

The versatility of the instrument has proven to be useful in numerous applications across different astronomical fields, beyond its initial application to transient astronomy. 
Given the larger sky coverage of ZTF over iPTF: 3760\,deg$^2$/h vs. 247\,deg$^2$/h, we expect the SEDM will be continuously over-subscribed once the survey becomes operational. 

Currently there are around 40 optical operating telescopes larger than 3\,m in the world. However, the number of smaller telescopes with diameters between 1$-$3\,m, is much larger, around 150. Hence, the possibility to replicate the instrument on other 1$-$3\,m class telescopes could solve the ``follow-up drought'', allowing new surveys to obtaining complete characterization of transient for a magnitude limited survey, providing unbiased statistical samples of numerous transient families in our nearby Universe.

\bibliographystyle{aasjournal}
\bibliography{references}

\begin{thebibliography}{}
\expandafter\ifx\csname natexlab\endcsname\relax\def\natexlab#1{#1}\fi

\bibitem[{{Baade} \& {Zwicky}(1934)}]{BaadeZwicky1934PNAS}
{Baade}, W., \& {Zwicky}, F. 1934, Proceedings of the National Academy of
  Science, 20, 254

\bibitem[{{Bellm}(2014)}]{Bellm2014htu}
{Bellm}, E. 2014, in The Third Hot-wiring the Transient Universe Workshop, ed.
  P.~R. {Wozniak}, M.~J. {Graham}, A.~A. {Mahabal}, \& R.~{Seaman}, 27--33

\bibitem[{{Bertin}(2011)}]{Bertin2011ASPC}
{Bertin}, E. 2011, in Astronomical Society of the Pacific Conference Series,
  Vol. 442, Astronomical Data Analysis Software and Systems XX, ed. I.~N.
  {Evans}, A.~{Accomazzi}, D.~J. {Mink}, \& A.~H. {Rots}, 435

\bibitem[{{Bertin} \& {Arnouts}(1996)}]{BertinArnouts1996AA}
{Bertin}, E., \& {Arnouts}, S. 1996, \aaps, 117, 393

\bibitem[{{Bertin} {et~al.}(2002){Bertin}, {Mellier}, {Radovich}, {Missonnier},
  {Didelon}, \& {Morin}}]{Bertin2002ASPC}
{Bertin}, E., {Mellier}, Y., {Radovich}, M., {et~al.} 2002, in Astronomical
  Society of the Pacific Conference Series, Vol. 281, Astronomical Data
  Analysis Software and Systems XI, ed. D.~A. {Bohlender}, D.~{Durand}, \&
  T.~H. {Handley}, 228

\bibitem[{{Blagorodnova} {et~al.}(2017){Blagorodnova}, {Gezari}, {Hung},
  {Kulkarni}, {Cenko}, {Pasham}, {Yan}, {Arcavi}, {Ben-Ami}, {Bue}, {Cantwell},
  {Cao}, {Castro-Tirado}, {Fender}, {Fremling}, {Gal-Yam}, {Ho}, {Horesh},
  {Hosseinzadeh}, {Kasliwal}, {Kong}, {Laher}, {Leloudas}, {Lunnan}, {Masci},
  {Mooley}, {Neill}, {Nugent}, {Powell}, {Valeev}, {Vreeswijk}, {Walters}, \&
  {Wozniak}}]{Blagorodnova2017ApJ}
{Blagorodnova}, N., {Gezari}, S., {Hung}, T., {et~al.} 2017, \apj, 844, 46

\bibitem[{{Burbidge} {et~al.}(1957){Burbidge}, {Burbidge}, {Fowler}, \&
  {Hoyle}}]{Burbidge1957RvMP}
{Burbidge}, E.~M., {Burbidge}, G.~R., {Fowler}, W.~A., \& {Hoyle}, F. 1957,
  Reviews of Modern Physics, 29, 547

\bibitem[{{Cenko} {et~al.}(2006){Cenko}, {Fox}, {Moon}, {Harrison}, {Kulkarni},
  {Henning}, {Guzman}, {Bonati}, {Smith}, {Thicksten}, {Doyle}, {Petrie},
  {Gal-Yam}, {Soderberg}, {Anagnostou}, \& {Laity}}]{Cenko2006PASP}
{Cenko}, S.~B., {Fox}, D.~B., {Moon}, D.-S., {et~al.} 2006, \pasp, 118, 1396

\bibitem[{{Drake} {et~al.}(2009){Drake}, {Djorgovski}, {Mahabal}, {Beshore},
  {Larson}, {Graham}, {Williams}, {Christensen}, {Catelan}, {Boattini},
  {Gibbs}, {Hill}, \& {Kowalski}}]{Drake2009ApJ}
{Drake}, A.~J., {Djorgovski}, S.~G., {Mahabal}, A., {et~al.} 2009, \apj, 696,
  870

\bibitem[{{Fremling} {et~al.}(2016){Fremling}, {Sollerman}, {Taddia}, {Ergon},
  {Fraser}, {Karamehmetoglu}, {Valenti}, {Jerkstrand}, {Arcavi}, {Bufano},
  {Elias Rosa}, {Filippenko}, {Fox}, {Gal-Yam}, {Howell}, {Kotak}, {Mazzali},
  {Milisavljevic}, {Nugent}, {Nyholm}, {Pian}, \&
  {Smartt}}]{2016A&A...593A..68F}
{Fremling}, C., {Sollerman}, J., {Taddia}, F., {et~al.} 2016, \aap, 593, A68

\bibitem[{{Fukugita} {et~al.}(1996){Fukugita}, {Ichikawa}, {Gunn}, {Doi},
  {Shimasaku}, \& {Schneider}}]{Fukugita1996}
{Fukugita}, M., {Ichikawa}, T., {Gunn}, J.~E., {et~al.} 1996, \aj, 111, 1748

\bibitem[{{Gal-Yam} {et~al.}(2008){Gal-Yam}, {Maoz}, {Guhathakurta}, \&
  {Filippenko}}]{Gal-Yam2008ApJ}
{Gal-Yam}, A., {Maoz}, D., {Guhathakurta}, P., \& {Filippenko}, A.~V. 2008,
  \apj, 680, 550

\bibitem[{{Gezari} {et~al.}(2017){Gezari}, {Hung}, {Cenko}, {Blagorodnova},
  {Yan}, {Kulkarni}, {Mooley}, {Kong}, {Cantwell}, {Yu}, {Cao}, {Fremling},
  {Neill}, {Ngeow}, {Nugent}, \& {Wozniak}}]{Gezari2017ApJ}
{Gezari}, S., {Hung}, T., {Cenko}, S.~B., {et~al.} 2017, \apj, 835, 144

\bibitem[{{Goobar} {et~al.}(2017){Goobar}, {Amanullah}, {Kulkarni}, {Nugent},
  {Johansson}, {Steidel}, {Law}, {M{\"o}rtsell}, {Quimby}, {Blagorodnova},
  {Brandeker}, {Cao}, {Cooray}, {Ferretti}, {Fremling}, {Hangard}, {Kasliwal},
  {Kupfer}, {Lunnan}, {Masci}, {Miller}, {Nayyeri}, {Neill}, {Ofek},
  {Papadogiannakis}, {Petrushevska}, {Ravi}, {Sollerman}, {Sullivan}, {Taddia},
  {Walters}, {Wilson}, {Yan}, \& {Yaron}}]{Goobar2017Sci}
{Goobar}, A., {Amanullah}, R., {Kulkarni}, S.~R., {et~al.} 2017, Science, 356,
  291

\bibitem[{Ivezi\'c {et~al.}(2008)Ivezi\'c, Tyson, Acosta, Allsman, Anderson,
  Andrew, Angel, Axelrod, Barr, Becker, {et~al.}}]{ivezic2008lsst}
Ivezi\'c, v., Tyson, J.~A., Acosta, E., {et~al.} 2008, arXiv:0805.2366v4

\bibitem[{{Kaiser} {et~al.}(2002){Kaiser}, {Aussel}, {Burke}, {Boesgaard},
  {Chambers}, {Chun}, {Heasley}, {Hodapp}, {Hunt}, {Jedicke}, {Jewitt},
  {Kudritzki}, {Luppino}, {Maberry}, {Magnier}, {Monet}, {Onaka}, {Pickles},
  {Rhoads}, {Simon}, {Szalay}, {Szapudi}, {Tholen}, {Tonry}, {Waterson}, \&
  {Wick}}]{Kaiser2002SPIE}
{Kaiser}, N., {Aussel}, H., {Burke}, B.~E., {et~al.} 2002, in \procspie, Vol.
  4836, Survey and Other Telescope Technologies and Discoveries, ed. J.~A.
  {Tyson} \& S.~{Wolff}, 154--164

\bibitem[{{Lang} {et~al.}(2010){Lang}, {Hogg}, {Mierle}, {Blanton}, \&
  {Roweis}}]{Lang2010}
{Lang}, D., {Hogg}, D.~W., {Mierle}, K., {Blanton}, M., \& {Roweis}, S. 2010,
  \aj, 139, 1782

\bibitem[{{Magnier} {et~al.}(2016){Magnier}, {Schlafly}, {Finkbeiner}, {Tonry},
  {Goldman}, {R{\"o}ser}, {Schilbach}, {Chambers}, {Flewelling}, {Huber},
  {Price}, {Sweeney}, {Waters}, {Denneau}, {Draper}, {Hodapp}, {Jedicke},
  {Kudritzki}, {Metcalfe}, {Stubbs}, \& {Wainscoast}}]{Magnier2016arXiv}
{Magnier}, E.~A., {Schlafly}, E.~F., {Finkbeiner}, D.~P., {et~al.} 2016, ArXiv
  e-prints, arXiv:1612.05242

\bibitem[{{Miller} {et~al.}(2017){Miller}, {Kasliwal}, {Cao}, {Goobar}, {Kne{\v
  z}evi{\'c}}, {Laher}, {Lunnan}, {Masci}, {Nugent}, {Perley}, {Petrushevska},
  {Quimby}, {Rebbapragada}, {.~Sollerman}, {Taddia}, \&
  {Kulkarni}}]{Miller2017arXiv}
{Miller}, A.~A., {Kasliwal}, M.~M., {Cao}, Y., {et~al.} 2017, ArXiv e-prints,
  arXiv:1703.07449

\bibitem[{NationalResearchCouncil(2010)}]{NAP12951}
NationalResearchCouncil, T. 2010, New Worlds, New Horizons in Astronomy and
  Astrophysics (Washington, DC: The National Academies Press),
  doi:10.17226/12951

\bibitem[{{Ofek}(2014)}]{Ofek2014ascl}
{Ofek}, E.~O. 2014, {MATLAB package for astronomy and astrophysics},
  Astrophysics Source Code Library, , , ascl:1407.005

\bibitem[{{Oke} \& {Gunn}(1983)}]{OkeGunn1983}
{Oke}, J.~B., \& {Gunn}, J.~E. 1983, \apj, 266, 713

\bibitem[{{Perlmutter} {et~al.}(1999){Perlmutter}, {Aldering}, {Goldhaber},
  {Knop}, {Nugent}, {Castro}, {Deustua}, {Fabbro}, {Goobar}, {Groom}, {Hook},
  {Kim}, {Kim}, {Lee}, {Nunes}, {Pain}, {Pennypacker}, {Quimby}, {Lidman},
  {Ellis}, {Irwin}, {McMahon}, {Ruiz-Lapuente}, {Walton}, {Schaefer}, {Boyle},
  {Filippenko}, {Matheson}, {Fruchter}, {Panagia}, {Newberg}, {Couch}, \&
  {Project}}]{Perlmutter1999ApJ}
{Perlmutter}, S., {Aldering}, G., {Goldhaber}, G., {et~al.} 1999, \apj, 517,
  565

\bibitem[{{Piascik} {et~al.}(2014){Piascik}, {Steele}, {Bates}, {Mottram},
  {Smith}, {Barnsley}, \& {Bolton}}]{Piascik2014SPIE}
{Piascik}, A.~S., {Steele}, I.~A., {Bates}, S.~D., {et~al.} 2014, in \procspie,
  Vol. 9147, Ground-based and Airborne Instrumentation for Astronomy V, 91478H

\bibitem[{{Rau} {et~al.}(2009){Rau}, {Kulkarni}, {Law}, {Bloom}, {Ciardi},
  {Djorgovski}, {Fox}, {Gal-Yam}, {Grillmair}, {Kasliwal}, {Nugent}, {Ofek},
  {Quimby}, {Reach}, {Shara}, {Bildsten}, {Cenko}, {Drake}, {Filippenko},
  {Helfand}, {Helou}, {Howell}, {Poznanski}, \& {Sullivan}}]{Rau2009PASP}
{Rau}, A., {Kulkarni}, S.~R., {Law}, N.~M., {et~al.} 2009, \pasp, 121, 1334

\bibitem[{{Riess} {et~al.}(1998){Riess}, {Filippenko}, {Challis},
  {Clocchiatti}, {Diercks}, {Garnavich}, {Gilliland}, {Hogan}, {Jha},
  {Kirshner}, {Leibundgut}, {Phillips}, {Reiss}, {Schmidt}, {Schommer},
  {Smith}, {Spyromilio}, {Stubbs}, {Suntzeff}, \& {Tonry}}]{Riess1998AJ}
{Riess}, A.~G., {Filippenko}, A.~V., {Challis}, P., {et~al.} 1998, \aj, 116,
  1009

\bibitem[{{Sand} {et~al.}(2011){Sand}, {Brown}, {Haynes}, \&
  {Dubberley}}]{Sand2011AAS}
{Sand}, D.~J., {Brown}, T., {Haynes}, R., \& {Dubberley}, M. 2011, in Bulletin
  of the American Astronomical Society, Vol.~43, American Astronomical Society
  Meeting Abstracts \#218, 132.03

\bibitem[{{Shappee} {et~al.}(2014){Shappee}, {Prieto}, {Grupe}, {Kochanek},
  {Stanek}, {De Rosa}, {Mathur}, {Zu}, {Peterson}, {Pogge}, {Komossa}, {Im},
  {Jencson}, {Holoien}, {Basu}, {Beacom}, {Szczygie{\l}}, {Brimacombe},
  {Adams}, {Campillay}, {Choi}, {Contreras}, {Dietrich}, {Dubberley},
  {Elphick}, {Foale}, {Giustini}, {Gonzalez}, {Hawkins}, {Howell}, {Hsiao},
  {Koss}, {Leighly}, {Morrell}, {Mudd}, {Mullins}, {Nugent}, {Parrent},
  {Phillips}, {Pojmanski}, {Rosing}, {Ross}, {Sand}, {Terndrup}, {Valenti},
  {Walker}, \& {Yoon}}]{Shappee2014ApJ}
{Shappee}, B.~J., {Prieto}, J.~L., {Grupe}, D., {et~al.} 2014, \apj, 788, 48

\bibitem[{{Tonry}(2011)}]{Tonry2011PASP}
{Tonry}, J.~L. 2011, \pasp, 123, 58

\bibitem[{{van Dokkum}(2001)}]{vanDokkum2001PASP}
{van Dokkum}, P.~G. 2001, \pasp, 113, 1420

\bibitem[{{Yu} {et~al.}(2016){Yu}, {Lin}, {Lin}, {Lee}, {Konidaris}, {Ngeow},
  {Ip}, {Chen}, {Chen}, {Malkan}, {Chang}, {Laher}, {Huang}, {Cheng},
  {Edelson}, {Ritter}, {Quimby}, {Ben-Ami}, {Ofek}, {Surace}, \&
  {Kulkarni}}]{Yu2016AJ}
{Yu}, P.-C., {Lin}, C.-C., {Lin}, H.-W., {et~al.} 2016, \aj, 151, 121

\end{thebibliography}

\facility{PO:1.2\,m, PO:1.5\,m  }

\software{ IRAF,  PYRAF, \texttt{PSFex} \citep{Bertin2011ASPC}, SExtractor \citep{BertinArnouts1996AA}, SWarp \citep{Bertin2002ASPC} }.

\section*{Acknowledgments\label{sec:acknowledgements}}
The SEDM was funded via NSF 1106171. 
This work was supported by the GROWTH project funded by the National Science Foundation under Grant No 1545949. 
The Oskar Klein Centre is funded by the Swedish Research Council.

\appendix

\begin{table*}
\centering
\begin{small}
\caption{List of spectroscopic standard stars routinely used for flux calibration by the IFU pipeline.}
\begin{tabular}{lcccc}
\hline
Name & R.A.& Decl. & V (mag)& SpecType \\ \hline
BD+25d4655  & 21:59:42.02  & +26:25:58.1  & 9.76  & O \\
BD+28d4211  & 21:51:11.07  & +28:51:51.8  & 10.51  & Op\\
BD+33d2642  & 15:51:59.86  & +32:56:54.8  & 10.81  & B2IV \\
BD+75d325   & 08:10:49.31  & +74:57:57.5  & 9.54  & O5p \\
Feige110*   & 23:19:58.39  & -05:09:55.8  & 11.82  & DOp \\
Feige34*    & 10:39:36.71  & +43:06:10.1  & 11.18  & DO \\
GD50		& 03:48:50.06 & $-$00:58:30.4 & 14.06 &DA2\\
HZ2  		& 04:12:43.51  & +11:51:50.4  & 13.86  & DA3 \\
HZ4			&  03:55:21.70 & +09:47:18.7 &  14.51& DA4\\
LB227		& 04:09:28.76 & +17:07:54.4 & 15.32 &DA4\\
\hline
\end{tabular}
\label{tab:stdlog}
\end{small}
\end{table*}

\end{document}